\documentclass[longauth]{aa}

\usepackage{graphicx}
\usepackage{xcolor}
\usepackage{txfonts}
\usepackage{placeins}
\usepackage{stfloats}
\usepackage{euclid}
\usepackage{lastpage}
\usepackage{float}
\defcitealias{Bisigello2025b}{B25}
\defcitealias{Gentile2024a}{G24}
\defcitealias{Gentile2025}{G25}

\usepackage{siunitx}
\DeclareSIUnit{\jansky}{Jy}
\DeclareSIUnit{\beam}{beam}
\usepackage[pdfencoding=auto,psdextra]{hyperref}
\hypersetup{
    colorlinks=true,
    linkcolor=blue,
    filecolor=magenta,      
    urlcolor=blue,
    citecolor=blue
}
\usepackage[switch, modulo]{lineno}
\linenumbers

\usepackage[nameinlink,capitalise]{cleveref}
\crefname{section}{Sect.}{Sects.}
\Crefname{section}{Section}{Sections}
\crefname{figure}{Fig.}{Figs.}
\Crefname{figure}{Figure}{Figures}
\crefname{equation}{Eq.}{Eqs.}
\Crefname{equation}{Equation}{Equations}
\crefname{table}{Table}{Tables}
\crefname{appendix}{Appendix}{Appendices}

\begin{document} 
\title{Euclid Quick Data Release (Q1)}
\subtitle{Searching for radio-selected \Euclid-dark galaxies in the EDF-N}   

\titlerunning{LOFAR-selected \Euclid-dark galaxies}

\newcommand{\orcid}[1]{} 		   
\author{Euclid Collaboration: M.~Giulietti\orcid{0000-0002-1847-4496}\thanks{\email{m.giulietti@ira.inaf.it}}\inst{\ref{aff1}}
\and I.~Prandoni\orcid{0000-0001-9680-7092}\inst{\ref{aff1}}
\and L.~Bisigello\orcid{0000-0003-0492-4924}\inst{\ref{aff2}}
\and G.~Rodighiero\orcid{0000-0002-9415-2296}\inst{\ref{aff3},\ref{aff2}}
\and M.~Talia\orcid{0000-0003-4352-2063}\inst{\ref{aff4},\ref{aff5}}
\and F.~Gentile\orcid{0000-0002-8008-9871}\inst{\ref{aff6},\ref{aff5}}
\and G.~Girardi\orcid{0009-0005-6156-4066}\inst{\ref{aff3},\ref{aff2}}
\and M.~Bondi\orcid{0000-0002-9553-7999}\inst{\ref{aff1}}
\and L.~Wang\orcid{0000-0002-6736-9158}\inst{\ref{aff7},\ref{aff8}}
\and A.~La~Marca\orcid{0000-0002-7217-5120}\inst{\ref{aff9}}
\and P.~A.~C.~Cunha\orcid{0000-0002-9454-859X}\inst{\ref{aff10}}
\and R.~Hill\orcid{0009-0008-8718-0644}\inst{\ref{aff11}}
\and A.~Abghari\orcid{0009-0003-2250-3880}\inst{\ref{aff11}}
\and D.~Scott\orcid{0000-0002-6878-9840}\inst{\ref{aff11}}
\and G.~Zamorani\orcid{0000-0002-2318-301X}\inst{\ref{aff5}}
\and G.~A.~Mamon\orcid{0000-0001-8956-5953}\inst{\ref{aff12},\ref{aff13}}
\and A.~Lapi\orcid{0000-0002-4882-1735}\inst{\ref{aff14},\ref{aff15},\ref{aff1},\ref{aff16}}
\and M.~Behiri\orcid{0000-0002-6444-8547}\inst{\ref{aff5}}
\and G.~Santhosh\orcid{0009-0008-8628-3537}\inst{\ref{aff1},\ref{aff4}}
\and H.~J.~A.~Rottgering\orcid{0000-0001-8887-2257}\inst{\ref{aff17}}
\and R.~Gilli\orcid{0000-0001-8121-6177}\inst{\ref{aff5}}
\and R.~Scaramella\orcid{0000-0003-2229-193X}\inst{\ref{aff18}}
\and G.~Gandolfi\orcid{0000-0003-3248-5666}\inst{\ref{aff18}}
\and S.~Andreon\orcid{0000-0002-2041-8784}\inst{\ref{aff19}}
\and N.~Auricchio\orcid{0000-0003-4444-8651}\inst{\ref{aff5}}
\and C.~Baccigalupi\orcid{0000-0002-8211-1630}\inst{\ref{aff15},\ref{aff20},\ref{aff16},\ref{aff14}}
\and M.~Baldi\orcid{0000-0003-4145-1943}\inst{\ref{aff10},\ref{aff5},\ref{aff21}}
\and A.~Balestra\orcid{0000-0002-6967-261X}\inst{\ref{aff2}}
\and S.~Bardelli\orcid{0000-0002-8900-0298}\inst{\ref{aff5}}
\and P.~Battaglia\orcid{0000-0002-7337-5909}\inst{\ref{aff5}}
\and A.~Biviano\orcid{0000-0002-0857-0732}\inst{\ref{aff20},\ref{aff15}}
\and M.~Bolzonella\orcid{0000-0003-3278-4607}\inst{\ref{aff5}}
\and E.~Branchini\orcid{0000-0002-0808-6908}\inst{\ref{aff22},\ref{aff23},\ref{aff19}}
\and M.~Brescia\orcid{0000-0001-9506-5680}\inst{\ref{aff24},\ref{aff25}}
\and J.~Brinchmann\orcid{0000-0003-4359-8797}\inst{\ref{aff26},\ref{aff27},\ref{aff28}}
\and S.~Camera\orcid{0000-0003-3399-3574}\inst{\ref{aff29},\ref{aff30},\ref{aff31}}
\and G.~Ca\~nas-Herrera\orcid{0000-0003-2796-2149}\inst{\ref{aff9},\ref{aff17}}
\and V.~Capobianco\orcid{0000-0002-3309-7692}\inst{\ref{aff31}}
\and C.~Carbone\orcid{0000-0003-0125-3563}\inst{\ref{aff32}}
\and J.~Carretero\orcid{0000-0002-3130-0204}\inst{\ref{aff33},\ref{aff34}}
\and S.~Casas\orcid{0000-0002-4751-5138}\inst{\ref{aff35},\ref{aff36}}
\and M.~Castellano\orcid{0000-0001-9875-8263}\inst{\ref{aff18}}
\and G.~Castignani\orcid{0000-0001-6831-0687}\inst{\ref{aff5}}
\and S.~Cavuoti\orcid{0000-0002-3787-4196}\inst{\ref{aff25},\ref{aff37}}
\and K.~C.~Chambers\orcid{0000-0001-6965-7789}\inst{\ref{aff38}}
\and A.~Cimatti\inst{\ref{aff39}}
\and C.~Colodro-Conde\inst{\ref{aff40}}
\and G.~Congedo\orcid{0000-0003-2508-0046}\inst{\ref{aff41}}
\and C.~J.~Conselice\orcid{0000-0003-1949-7638}\inst{\ref{aff42}}
\and L.~Conversi\orcid{0000-0002-6710-8476}\inst{\ref{aff43},\ref{aff44}}
\and Y.~Copin\orcid{0000-0002-5317-7518}\inst{\ref{aff45}}
\and A.~Costille\inst{\ref{aff46}}
\and F.~Courbin\orcid{0000-0003-0758-6510}\inst{\ref{aff47},\ref{aff48},\ref{aff49}}
\and H.~M.~Courtois\orcid{0000-0003-0509-1776}\inst{\ref{aff50}}
\and M.~Cropper\orcid{0000-0003-4571-9468}\inst{\ref{aff51}}
\and A.~Da~Silva\orcid{0000-0002-6385-1609}\inst{\ref{aff52},\ref{aff53}}
\and H.~Degaudenzi\orcid{0000-0002-5887-6799}\inst{\ref{aff54}}
\and G.~De~Lucia\orcid{0000-0002-6220-9104}\inst{\ref{aff20}}
\and C.~Dolding\orcid{0009-0003-7199-6108}\inst{\ref{aff51}}
\and H.~Dole\orcid{0000-0002-9767-3839}\inst{\ref{aff55}}
\and F.~Dubath\orcid{0000-0002-6533-2810}\inst{\ref{aff54}}
\and C.~A.~J.~Duncan\orcid{0009-0003-3573-0791}\inst{\ref{aff41}}
\and X.~Dupac\inst{\ref{aff44}}
\and S.~Dusini\orcid{0000-0002-1128-0664}\inst{\ref{aff56}}
\and S.~Escoffier\orcid{0000-0002-2847-7498}\inst{\ref{aff57}}
\and M.~Farina\orcid{0000-0002-3089-7846}\inst{\ref{aff58}}
\and R.~Farinelli\inst{\ref{aff5}}
\and F.~Faustini\orcid{0000-0001-6274-5145}\inst{\ref{aff18},\ref{aff59}}
\and S.~Ferriol\inst{\ref{aff45}}
\and F.~Finelli\orcid{0000-0002-6694-3269}\inst{\ref{aff5},\ref{aff60}}
\and S.~Fotopoulou\orcid{0000-0002-9686-254X}\inst{\ref{aff61}}
\and M.~Frailis\orcid{0000-0002-7400-2135}\inst{\ref{aff20}}
\and E.~Franceschi\orcid{0000-0002-0585-6591}\inst{\ref{aff5}}
\and M.~Fumana\orcid{0000-0001-6787-5950}\inst{\ref{aff32}}
\and L.~Gabarra\orcid{0000-0002-8486-8856}\inst{\ref{aff62}}
\and S.~Galeotta\orcid{0000-0002-3748-5115}\inst{\ref{aff20}}
\and K.~George\orcid{0000-0002-1734-8455}\inst{\ref{aff63}}
\and B.~Gillis\orcid{0000-0002-4478-1270}\inst{\ref{aff41}}
\and C.~Giocoli\orcid{0000-0002-9590-7961}\inst{\ref{aff5},\ref{aff21}}
\and J.~Gracia-Carpio\inst{\ref{aff64}}
\and A.~Grazian\orcid{0000-0002-5688-0663}\inst{\ref{aff2}}
\and F.~Grupp\inst{\ref{aff64},\ref{aff65}}
\and S.~Gwyn\orcid{0000-0001-8221-8406}\inst{\ref{aff66}}
\and W.~G.~Hartley\inst{\ref{aff54}}
\and S.~V.~H.~Haugan\orcid{0000-0001-9648-7260}\inst{\ref{aff67}}
\and S.~Hemmati\orcid{0000-0003-2226-5395}\inst{\ref{aff68}}
\and J.~Hoar\inst{\ref{aff44}}
\and W.~Holmes\inst{\ref{aff69}}
\and A.~Hornstrup\orcid{0000-0002-3363-0936}\inst{\ref{aff70},\ref{aff71}}
\and M.~Huertas-Company\orcid{0000-0002-1416-8483}\inst{\ref{aff40},\ref{aff72},\ref{aff73},\ref{aff74}}
\and K.~Jahnke\orcid{0000-0003-3804-2137}\inst{\ref{aff75}}
\and M.~Jhabvala\inst{\ref{aff76}}
\and B.~Joachimi\orcid{0000-0001-7494-1303}\inst{\ref{aff77}}
\and E.~Keih\"anen\orcid{0000-0003-1804-7715}\inst{\ref{aff78}}
\and S.~Kermiche\orcid{0000-0002-0302-5735}\inst{\ref{aff57}}
\and A.~Kiessling\orcid{0000-0002-2590-1273}\inst{\ref{aff69}}
\and M.~Kilbinger\orcid{0000-0001-9513-7138}\inst{\ref{aff79}}
\and B.~Kubik\orcid{0009-0006-5823-4880}\inst{\ref{aff45}}
\and M.~K\"ummel\orcid{0000-0003-2791-2117}\inst{\ref{aff65}}
\and M.~Kunz\orcid{0000-0002-3052-7394}\inst{\ref{aff80}}
\and H.~Kurki-Suonio\orcid{0000-0002-4618-3063}\inst{\ref{aff81},\ref{aff82}}
\and A.~M.~C.~Le~Brun\orcid{0000-0002-0936-4594}\inst{\ref{aff83}}
\and S.~Ligori\orcid{0000-0003-4172-4606}\inst{\ref{aff31}}
\and P.~B.~Lilje\orcid{0000-0003-4324-7794}\inst{\ref{aff67}}
\and V.~Lindholm\orcid{0000-0003-2317-5471}\inst{\ref{aff81},\ref{aff82}}
\and I.~Lloro\orcid{0000-0001-5966-1434}\inst{\ref{aff84}}
\and M.~Magliocchetti\orcid{0000-0001-9158-4838}\inst{\ref{aff58}}
\and G.~Mainetti\orcid{0000-0003-2384-2377}\inst{\ref{aff85}}
\and D.~Maino\inst{\ref{aff86},\ref{aff32},\ref{aff87}}
\and E.~Maiorano\orcid{0000-0003-2593-4355}\inst{\ref{aff5}}
\and O.~Mansutti\orcid{0000-0001-5758-4658}\inst{\ref{aff20}}
\and S.~Marcin\inst{\ref{aff88}}
\and O.~Marggraf\orcid{0000-0001-7242-3852}\inst{\ref{aff89}}
\and M.~Martinelli\orcid{0000-0002-6943-7732}\inst{\ref{aff18},\ref{aff90}}
\and N.~Martinet\orcid{0000-0003-2786-7790}\inst{\ref{aff46}}
\and F.~Marulli\orcid{0000-0002-8850-0303}\inst{\ref{aff4},\ref{aff5},\ref{aff21}}
\and R.~J.~Massey\orcid{0000-0002-6085-3780}\inst{\ref{aff91}}
\and N.~Mauri\orcid{0000-0001-8196-1548}\inst{\ref{aff39},\ref{aff21}}
\and E.~Medinaceli\orcid{0000-0002-4040-7783}\inst{\ref{aff5}}
\and S.~Mei\orcid{0000-0002-2849-559X}\inst{\ref{aff92},\ref{aff93}}
\and Y.~Mellier\thanks{Deceased}\inst{\ref{aff13},\ref{aff12}}
\and M.~Meneghetti\orcid{0000-0003-1225-7084}\inst{\ref{aff5},\ref{aff21}}
\and E.~Merlin\orcid{0000-0001-6870-8900}\inst{\ref{aff18}}
\and G.~Meylan\inst{\ref{aff94}}
\and P.~Monaco\orcid{0000-0003-2083-7564}\inst{\ref{aff95},\ref{aff20},\ref{aff16},\ref{aff15}}
\and A.~Mora\orcid{0000-0002-1922-8529}\inst{\ref{aff96}}
\and M.~Moresco\orcid{0000-0002-7616-7136}\inst{\ref{aff4},\ref{aff5}}
\and C.~Moretti\orcid{0000-0003-3314-8936}\inst{\ref{aff20},\ref{aff15},\ref{aff16},\ref{aff14}}
\and L.~Moscardini\orcid{0000-0002-3473-6716}\inst{\ref{aff4},\ref{aff5},\ref{aff21}}
\and R.~Nakajima\orcid{0009-0009-1213-7040}\inst{\ref{aff89}}
\and C.~Neissner\orcid{0000-0001-8524-4968}\inst{\ref{aff97},\ref{aff34}}
\and R.~C.~Nichol\orcid{0000-0003-0939-6518}\inst{\ref{aff98}}
\and S.-M.~Niemi\orcid{0009-0005-0247-0086}\inst{\ref{aff9}}
\and C.~Padilla\orcid{0000-0001-7951-0166}\inst{\ref{aff97}}
\and S.~Paltani\orcid{0000-0002-8108-9179}\inst{\ref{aff54}}
\and F.~Pasian\orcid{0000-0002-4869-3227}\inst{\ref{aff20}}
\and K.~Pedersen\inst{\ref{aff99}}
\and W.~J.~Percival\orcid{0000-0002-0644-5727}\inst{\ref{aff100},\ref{aff101},\ref{aff102}}
\and V.~Pettorino\orcid{0000-0002-4203-9320}\inst{\ref{aff9}}
\and A.~Pezzotta\orcid{0000-0003-0726-2268}\inst{\ref{aff19}}
\and S.~Pires\orcid{0000-0002-0249-2104}\inst{\ref{aff79}}
\and G.~Polenta\orcid{0000-0003-4067-9196}\inst{\ref{aff59}}
\and M.~Poncet\inst{\ref{aff103}}
\and L.~A.~Popa\inst{\ref{aff104}}
\and C.~Porciani\orcid{0000-0002-7797-2508}\inst{\ref{aff89}}
\and L.~Pozzetti\orcid{0000-0001-7085-0412}\inst{\ref{aff5}}
\and F.~Raison\orcid{0000-0002-7819-6918}\inst{\ref{aff64}}
\and A.~Renzi\orcid{0000-0001-9856-1970}\inst{\ref{aff3},\ref{aff56}}
\and J.~Rhodes\orcid{0000-0002-4485-8549}\inst{\ref{aff69}}
\and G.~Riccio\inst{\ref{aff25}}
\and I.~Risso\orcid{0000-0003-2525-7761}\inst{\ref{aff19},\ref{aff23}}
\and E.~Romelli\orcid{0000-0003-3069-9222}\inst{\ref{aff20}}
\and M.~Roncarelli\orcid{0000-0001-9587-7822}\inst{\ref{aff5}}
\and R.~Saglia\orcid{0000-0003-0378-7032}\inst{\ref{aff65},\ref{aff64}}
\and Z.~Sakr\orcid{0000-0002-4823-3757}\inst{\ref{aff105},\ref{aff106},\ref{aff107}}
\and D.~Sapone\orcid{0000-0001-7089-4503}\inst{\ref{aff108}}
\and B.~Sartoris\orcid{0000-0003-1337-5269}\inst{\ref{aff65},\ref{aff20}}
\and J.~A.~Schewtschenko\orcid{0000-0002-4913-6393}\inst{\ref{aff41}}
\and M.~Schirmer\orcid{0000-0003-2568-9994}\inst{\ref{aff75}}
\and P.~Schneider\orcid{0000-0001-8561-2679}\inst{\ref{aff89}}
\and T.~Schrabback\orcid{0000-0002-6987-7834}\inst{\ref{aff109}}
\and A.~Secroun\orcid{0000-0003-0505-3710}\inst{\ref{aff57}}
\and G.~Seidel\orcid{0000-0003-2907-353X}\inst{\ref{aff75}}
\and S.~Serrano\orcid{0000-0002-0211-2861}\inst{\ref{aff110},\ref{aff111},\ref{aff112}}
\and P.~Simon\inst{\ref{aff89}}
\and C.~Sirignano\orcid{0000-0002-0995-7146}\inst{\ref{aff3},\ref{aff56}}
\and G.~Sirri\orcid{0000-0003-2626-2853}\inst{\ref{aff21}}
\and L.~Stanco\orcid{0000-0002-9706-5104}\inst{\ref{aff56}}
\and J.-L.~Starck\orcid{0000-0003-2177-7794}\inst{\ref{aff79}}
\and J.~Steinwagner\orcid{0000-0001-7443-1047}\inst{\ref{aff64}}
\and P.~Tallada-Cresp\'{i}\orcid{0000-0002-1336-8328}\inst{\ref{aff33},\ref{aff34}}
\and A.~N.~Taylor\inst{\ref{aff41}}
\and H.~I.~Teplitz\orcid{0000-0002-7064-5424}\inst{\ref{aff113}}
\and I.~Tereno\orcid{0000-0002-4537-6218}\inst{\ref{aff52},\ref{aff114}}
\and N.~Tessore\orcid{0000-0002-9696-7931}\inst{\ref{aff77},\ref{aff51}}
\and S.~Toft\orcid{0000-0003-3631-7176}\inst{\ref{aff115},\ref{aff116}}
\and R.~Toledo-Moreo\orcid{0000-0002-2997-4859}\inst{\ref{aff117}}
\and F.~Torradeflot\orcid{0000-0003-1160-1517}\inst{\ref{aff34},\ref{aff33}}
\and I.~Tutusaus\orcid{0000-0002-3199-0399}\inst{\ref{aff112},\ref{aff110},\ref{aff106}}
\and L.~Valenziano\orcid{0000-0002-1170-0104}\inst{\ref{aff5},\ref{aff60}}
\and J.~Valiviita\orcid{0000-0001-6225-3693}\inst{\ref{aff81},\ref{aff82}}
\and T.~Vassallo\orcid{0000-0001-6512-6358}\inst{\ref{aff20}}
\and A.~Veropalumbo\orcid{0000-0003-2387-1194}\inst{\ref{aff19},\ref{aff23},\ref{aff22}}
\and Y.~Wang\orcid{0000-0002-4749-2984}\inst{\ref{aff113}}
\and J.~Weller\orcid{0000-0002-8282-2010}\inst{\ref{aff65},\ref{aff64}}
\and A.~Zacchei\orcid{0000-0003-0396-1192}\inst{\ref{aff20},\ref{aff15}}
\and E.~Zucca\orcid{0000-0002-5845-8132}\inst{\ref{aff5}}
\and V.~Allevato\orcid{0000-0001-7232-5152}\inst{\ref{aff25}}
\and M.~Ballardini\orcid{0000-0003-4481-3559}\inst{\ref{aff118},\ref{aff119},\ref{aff5}}
\and E.~Bozzo\orcid{0000-0002-8201-1525}\inst{\ref{aff54}}
\and C.~Burigana\orcid{0000-0002-3005-5796}\inst{\ref{aff1},\ref{aff60}}
\and R.~Cabanac\orcid{0000-0001-6679-2600}\inst{\ref{aff106}}
\and M.~Calabrese\orcid{0000-0002-2637-2422}\inst{\ref{aff120},\ref{aff32}}
\and A.~Cappi\inst{\ref{aff121},\ref{aff5}}
\and J.~A.~Escartin~Vigo\inst{\ref{aff64}}
\and R.~Maoli\orcid{0000-0002-6065-3025}\inst{\ref{aff122},\ref{aff18}}
\and J.~Mart\'{i}n-Fleitas\orcid{0000-0002-8594-569X}\inst{\ref{aff123}}
\and S.~Matthew\orcid{0000-0001-8448-1697}\inst{\ref{aff41}}
\and M.~Maturi\orcid{0000-0002-3517-2422}\inst{\ref{aff105},\ref{aff124}}
\and R.~B.~Metcalf\orcid{0000-0003-3167-2574}\inst{\ref{aff4},\ref{aff5}}
\and M.~P\"ontinen\orcid{0000-0001-5442-2530}\inst{\ref{aff81}}
\and V.~Scottez\orcid{0009-0008-3864-940X}\inst{\ref{aff13},\ref{aff125}}
\and M.~Sereno\orcid{0000-0003-0302-0325}\inst{\ref{aff5},\ref{aff21}}
\and M.~Tenti\orcid{0000-0002-4254-5901}\inst{\ref{aff21}}
\and M.~Viel\orcid{0000-0002-2642-5707}\inst{\ref{aff15},\ref{aff20},\ref{aff14},\ref{aff16},\ref{aff126}}
\and M.~Wiesmann\orcid{0009-0000-8199-5860}\inst{\ref{aff67}}
\and Y.~Akrami\orcid{0000-0002-2407-7956}\inst{\ref{aff127},\ref{aff128}}
\and I.~T.~Andika\orcid{0000-0001-6102-9526}\inst{\ref{aff129},\ref{aff130}}
\and S.~Anselmi\orcid{0000-0002-3579-9583}\inst{\ref{aff56},\ref{aff3},\ref{aff131}}
\and M.~Archidiacono\orcid{0000-0003-4952-9012}\inst{\ref{aff86},\ref{aff87}}
\and F.~Atrio-Barandela\orcid{0000-0002-2130-2513}\inst{\ref{aff132}}
\and D.~Bertacca\orcid{0000-0002-2490-7139}\inst{\ref{aff3},\ref{aff2},\ref{aff56}}
\and M.~Bethermin\orcid{0000-0002-3915-2015}\inst{\ref{aff133}}
\and A.~Blanchard\orcid{0000-0001-8555-9003}\inst{\ref{aff106}}
\and L.~Blot\orcid{0000-0002-9622-7167}\inst{\ref{aff134},\ref{aff83}}
\and M.~Bonici\orcid{0000-0002-8430-126X}\inst{\ref{aff100},\ref{aff32}}
\and S.~Borgani\orcid{0000-0001-6151-6439}\inst{\ref{aff95},\ref{aff15},\ref{aff20},\ref{aff16},\ref{aff126}}
\and M.~L.~Brown\orcid{0000-0002-0370-8077}\inst{\ref{aff42}}
\and S.~Bruton\orcid{0000-0002-6503-5218}\inst{\ref{aff135}}
\and A.~Calabro\orcid{0000-0003-2536-1614}\inst{\ref{aff18}}
\and B.~Camacho~Quevedo\orcid{0000-0002-8789-4232}\inst{\ref{aff15},\ref{aff14},\ref{aff20}}
\and F.~Caro\inst{\ref{aff18}}
\and C.~S.~Carvalho\inst{\ref{aff114}}
\and T.~Castro\orcid{0000-0002-6292-3228}\inst{\ref{aff20},\ref{aff16},\ref{aff15},\ref{aff126}}
\and F.~Cogato\orcid{0000-0003-4632-6113}\inst{\ref{aff4},\ref{aff5}}
\and S.~Conseil\orcid{0000-0002-3657-4191}\inst{\ref{aff45}}
\and A.~R.~Cooray\orcid{0000-0002-3892-0190}\inst{\ref{aff136}}
\and O.~Cucciati\orcid{0000-0002-9336-7551}\inst{\ref{aff5}}
\and S.~Davini\orcid{0000-0003-3269-1718}\inst{\ref{aff23}}
\and G.~Desprez\orcid{0000-0001-8325-1742}\inst{\ref{aff8}}
\and A.~D\'iaz-S\'anchez\orcid{0000-0003-0748-4768}\inst{\ref{aff137}}
\and J.~J.~Diaz\orcid{0000-0003-2101-1078}\inst{\ref{aff40}}
\and S.~Di~Domizio\orcid{0000-0003-2863-5895}\inst{\ref{aff22},\ref{aff23}}
\and J.~M.~Diego\orcid{0000-0001-9065-3926}\inst{\ref{aff138}}
\and M.~Y.~Elkhashab\orcid{0000-0001-9306-2603}\inst{\ref{aff20},\ref{aff16},\ref{aff95},\ref{aff15}}
\and A.~Enia\orcid{0000-0002-0200-2857}\inst{\ref{aff5}}
\and Y.~Fang\orcid{0000-0002-0334-6950}\inst{\ref{aff65}}
\and A.~Finoguenov\orcid{0000-0002-4606-5403}\inst{\ref{aff81}}
\and A.~Franco\orcid{0000-0002-4761-366X}\inst{\ref{aff139},\ref{aff140},\ref{aff141}}
\and K.~Ganga\orcid{0000-0001-8159-8208}\inst{\ref{aff92}}
\and J.~Garc\'ia-Bellido\orcid{0000-0002-9370-8360}\inst{\ref{aff127}}
\and T.~Gasparetto\orcid{0000-0002-7913-4866}\inst{\ref{aff18}}
\and V.~Gautard\inst{\ref{aff6}}
\and E.~Gaztanaga\orcid{0000-0001-9632-0815}\inst{\ref{aff112},\ref{aff110},\ref{aff142}}
\and F.~Giacomini\orcid{0000-0002-3129-2814}\inst{\ref{aff21}}
\and F.~Gianotti\orcid{0000-0003-4666-119X}\inst{\ref{aff5}}
\and G.~Gozaliasl\orcid{0000-0002-0236-919X}\inst{\ref{aff143},\ref{aff81}}
\and M.~Guidi\orcid{0000-0001-9408-1101}\inst{\ref{aff10},\ref{aff5}}
\and C.~M.~Gutierrez\orcid{0000-0001-7854-783X}\inst{\ref{aff72}}
\and A.~Hall\orcid{0000-0002-3139-8651}\inst{\ref{aff41}}
\and C.~Hern\'andez-Monteagudo\orcid{0000-0001-5471-9166}\inst{\ref{aff144},\ref{aff40}}
\and H.~Hildebrandt\orcid{0000-0002-9814-3338}\inst{\ref{aff145}}
\and J.~Hjorth\orcid{0000-0002-4571-2306}\inst{\ref{aff99}}
\and J.~J.~E.~Kajava\orcid{0000-0002-3010-8333}\inst{\ref{aff146},\ref{aff147}}
\and Y.~Kang\orcid{0009-0000-8588-7250}\inst{\ref{aff54}}
\and V.~Kansal\orcid{0000-0002-4008-6078}\inst{\ref{aff148},\ref{aff149}}
\and D.~Karagiannis\orcid{0000-0002-4927-0816}\inst{\ref{aff118},\ref{aff150}}
\and K.~Kiiveri\inst{\ref{aff78}}
\and J.~Kim\orcid{0000-0003-2776-2761}\inst{\ref{aff62}}
\and C.~C.~Kirkpatrick\inst{\ref{aff78}}
\and S.~Kruk\orcid{0000-0001-8010-8879}\inst{\ref{aff44}}
\and L.~Legrand\orcid{0000-0003-0610-5252}\inst{\ref{aff151},\ref{aff152}}
\and M.~Lembo\orcid{0000-0002-5271-5070}\inst{\ref{aff12},\ref{aff118},\ref{aff119}}
\and F.~Lepori\orcid{0009-0000-5061-7138}\inst{\ref{aff153}}
\and G.~F.~Lesci\orcid{0000-0002-4607-2830}\inst{\ref{aff4},\ref{aff5}}
\and J.~Lesgourgues\orcid{0000-0001-7627-353X}\inst{\ref{aff35}}
\and L.~Leuzzi\orcid{0009-0006-4479-7017}\inst{\ref{aff5}}
\and T.~I.~Liaudat\orcid{0000-0002-9104-314X}\inst{\ref{aff154}}
\and A.~Loureiro\orcid{0000-0002-4371-0876}\inst{\ref{aff155},\ref{aff156}}
\and J.~Macias-Perez\orcid{0000-0002-5385-2763}\inst{\ref{aff157}}
\and F.~Mannucci\orcid{0000-0002-4803-2381}\inst{\ref{aff158}}
\and C.~J.~A.~P.~Martins\orcid{0000-0002-4886-9261}\inst{\ref{aff159},\ref{aff26}}
\and L.~Maurin\orcid{0000-0002-8406-0857}\inst{\ref{aff55}}
\and M.~Miluzio\inst{\ref{aff44},\ref{aff160}}
\and G.~Morgante\inst{\ref{aff5}}
\and K.~Naidoo\orcid{0000-0002-9182-1802}\inst{\ref{aff142},\ref{aff77}}
\and P.~Natoli\orcid{0000-0003-0126-9100}\inst{\ref{aff118},\ref{aff119}}
\and A.~Navarro-Alsina\orcid{0000-0002-3173-2592}\inst{\ref{aff89}}
\and S.~Nesseris\orcid{0000-0002-0567-0324}\inst{\ref{aff127}}
\and D.~Paoletti\orcid{0000-0003-4761-6147}\inst{\ref{aff5},\ref{aff60}}
\and F.~Passalacqua\orcid{0000-0002-8606-4093}\inst{\ref{aff3},\ref{aff56}}
\and K.~Paterson\orcid{0000-0001-8340-3486}\inst{\ref{aff75}}
\and L.~Patrizii\inst{\ref{aff21}}
\and A.~Pisani\orcid{0000-0002-6146-4437}\inst{\ref{aff57}}
\and D.~Potter\orcid{0000-0002-0757-5195}\inst{\ref{aff153}}
\and S.~Quai\orcid{0000-0002-0449-8163}\inst{\ref{aff4},\ref{aff5}}
\and M.~Radovich\orcid{0000-0002-3585-866X}\inst{\ref{aff2}}
\and S.~Sacquegna\orcid{0000-0002-8433-6630}\inst{\ref{aff161}}
\and M.~Sahl\'en\orcid{0000-0003-0973-4804}\inst{\ref{aff162}}
\and D.~B.~Sanders\orcid{0000-0002-1233-9998}\inst{\ref{aff38}}
\and E.~Sarpa\orcid{0000-0002-1256-655X}\inst{\ref{aff14},\ref{aff126},\ref{aff16}}
\and A.~Schneider\orcid{0000-0001-7055-8104}\inst{\ref{aff153}}
\and D.~Sciotti\orcid{0009-0008-4519-2620}\inst{\ref{aff18},\ref{aff90}}
\and E.~Sellentin\inst{\ref{aff163},\ref{aff17}}
\and F.~Shankar\orcid{0000-0001-8973-5051}\inst{\ref{aff164}}
\and A.~Shulevski\orcid{0000-0002-1827-0469}\inst{\ref{aff165},\ref{aff8},\ref{aff17},\ref{aff166}}
\and L.~C.~Smith\orcid{0000-0002-3259-2771}\inst{\ref{aff167}}
\and J.~G.~Sorce\orcid{0000-0002-2307-2432}\inst{\ref{aff168},\ref{aff55}}
\and K.~Tanidis\orcid{0000-0001-9843-5130}\inst{\ref{aff62}}
\and C.~Tao\orcid{0000-0001-7961-8177}\inst{\ref{aff57}}
\and G.~Testera\inst{\ref{aff23}}
\and R.~Teyssier\orcid{0000-0001-7689-0933}\inst{\ref{aff169}}
\and S.~Tosi\orcid{0000-0002-7275-9193}\inst{\ref{aff22},\ref{aff23},\ref{aff19}}
\and A.~Troja\orcid{0000-0003-0239-4595}\inst{\ref{aff3},\ref{aff56}}
\and M.~Tucci\inst{\ref{aff54}}
\and A.~Venhola\orcid{0000-0001-6071-4564}\inst{\ref{aff170}}
\and D.~Vergani\orcid{0000-0003-0898-2216}\inst{\ref{aff5}}
\and G.~Verza\orcid{0000-0002-1886-8348}$^{\star\star}$\inst{\ref{aff171}}
\and P.~Vielzeuf\orcid{0000-0003-2035-9339}\inst{\ref{aff57}}
\and N.~A.~Walton\orcid{0000-0003-3983-8778}\inst{\ref{aff167}}}

\institute{INAF, Istituto di Radioastronomia, Via Piero Gobetti 101, 40129 Bologna, Italy\label{aff1}
\and
INAF-Osservatorio Astronomico di Padova, Via dell'Osservatorio 5, 35122 Padova, Italy\label{aff2}
\and
Dipartimento di Fisica e Astronomia "G. Galilei", Universit\`a di Padova, Via Marzolo 8, 35131 Padova, Italy\label{aff3}
\and
Dipartimento di Fisica e Astronomia "Augusto Righi" - Alma Mater Studiorum Universit\`a di Bologna, via Piero Gobetti 93/2, 40129 Bologna, Italy\label{aff4}
\and
INAF-Osservatorio di Astrofisica e Scienza dello Spazio di Bologna, Via Piero Gobetti 93/3, 40129 Bologna, Italy\label{aff5}
\and
CEA Saclay, DFR/IRFU, Service d'Astrophysique, Bat. 709, 91191 Gif-sur-Yvette, France\label{aff6}
\and
SRON Netherlands Institute for Space Research, Landleven 12, 9747 AD, Groningen, The Netherlands\label{aff7}
\and
Kapteyn Astronomical Institute, University of Groningen, PO Box 800, 9700 AV Groningen, The Netherlands\label{aff8}
\and
European Space Agency/ESTEC, Keplerlaan 1, 2201 AZ Noordwijk, The Netherlands\label{aff9}
\and
Dipartimento di Fisica e Astronomia, Universit\`a di Bologna, Via Gobetti 93/2, 40129 Bologna, Italy\label{aff10}
\and
Department of Physics and Astronomy, University of British Columbia, Vancouver, BC V6T 1Z1, Canada\label{aff11}
\and
Institut d'Astrophysique de Paris, UMR 7095, CNRS, and Sorbonne Universit\'e, 98 bis boulevard Arago, 75014 Paris, France\label{aff12}
\and
Institut d'Astrophysique de Paris, 98bis Boulevard Arago, 75014, Paris, France\label{aff13}
\and
SISSA, International School for Advanced Studies, Via Bonomea 265, 34136 Trieste TS, Italy\label{aff14}
\and
IFPU, Institute for Fundamental Physics of the Universe, via Beirut 2, 34151 Trieste, Italy\label{aff15}
\and
INFN, Sezione di Trieste, Via Valerio 2, 34127 Trieste TS, Italy\label{aff16}
\and
Leiden Observatory, Leiden University, Einsteinweg 55, 2333 CC Leiden, The Netherlands\label{aff17}
\and
INAF-Osservatorio Astronomico di Roma, Via Frascati 33, 00078 Monteporzio Catone, Italy\label{aff18}
\and
INAF-Osservatorio Astronomico di Brera, Via Brera 28, 20122 Milano, Italy\label{aff19}
\and
INAF-Osservatorio Astronomico di Trieste, Via G. B. Tiepolo 11, 34143 Trieste, Italy\label{aff20}
\and
INFN-Sezione di Bologna, Viale Berti Pichat 6/2, 40127 Bologna, Italy\label{aff21}
\and
Dipartimento di Fisica, Universit\`a di Genova, Via Dodecaneso 33, 16146, Genova, Italy\label{aff22}
\and
INFN-Sezione di Genova, Via Dodecaneso 33, 16146, Genova, Italy\label{aff23}
\and
Department of Physics "E. Pancini", University Federico II, Via Cinthia 6, 80126, Napoli, Italy\label{aff24}
\and
INAF-Osservatorio Astronomico di Capodimonte, Via Moiariello 16, 80131 Napoli, Italy\label{aff25}
\and
Instituto de Astrof\'isica e Ci\^encias do Espa\c{c}o, Universidade do Porto, CAUP, Rua das Estrelas, PT4150-762 Porto, Portugal\label{aff26}
\and
Faculdade de Ci\^encias da Universidade do Porto, Rua do Campo de Alegre, 4150-007 Porto, Portugal\label{aff27}
\and
European Southern Observatory, Karl-Schwarzschild-Str.~2, 85748 Garching, Germany\label{aff28}
\and
Dipartimento di Fisica, Universit\`a degli Studi di Torino, Via P. Giuria 1, 10125 Torino, Italy\label{aff29}
\and
INFN-Sezione di Torino, Via P. Giuria 1, 10125 Torino, Italy\label{aff30}
\and
INAF-Osservatorio Astrofisico di Torino, Via Osservatorio 20, 10025 Pino Torinese (TO), Italy\label{aff31}
\and
INAF-IASF Milano, Via Alfonso Corti 12, 20133 Milano, Italy\label{aff32}
\and
Centro de Investigaciones Energ\'eticas, Medioambientales y Tecnol\'ogicas (CIEMAT), Avenida Complutense 40, 28040 Madrid, Spain\label{aff33}
\and
Port d'Informaci\'{o} Cient\'{i}fica, Campus UAB, C. Albareda s/n, 08193 Bellaterra (Barcelona), Spain\label{aff34}
\and
Institute for Theoretical Particle Physics and Cosmology (TTK), RWTH Aachen University, 52056 Aachen, Germany\label{aff35}
\and
Deutsches Zentrum f\"ur Luft- und Raumfahrt e. V. (DLR), Linder H\"ohe, 51147 K\"oln, Germany\label{aff36}
\and
INFN section of Naples, Via Cinthia 6, 80126, Napoli, Italy\label{aff37}
\and
Institute for Astronomy, University of Hawaii, 2680 Woodlawn Drive, Honolulu, HI 96822, USA\label{aff38}
\and
Dipartimento di Fisica e Astronomia "Augusto Righi" - Alma Mater Studiorum Universit\`a di Bologna, Viale Berti Pichat 6/2, 40127 Bologna, Italy\label{aff39}
\and
Instituto de Astrof\'{\i}sica de Canarias, E-38205 La Laguna, Tenerife, Spain\label{aff40}
\and
Institute for Astronomy, University of Edinburgh, Royal Observatory, Blackford Hill, Edinburgh EH9 3HJ, UK\label{aff41}
\and
Jodrell Bank Centre for Astrophysics, Department of Physics and Astronomy, University of Manchester, Oxford Road, Manchester M13 9PL, UK\label{aff42}
\and
European Space Agency/ESRIN, Largo Galileo Galilei 1, 00044 Frascati, Roma, Italy\label{aff43}
\and
ESAC/ESA, Camino Bajo del Castillo, s/n., Urb. Villafranca del Castillo, 28692 Villanueva de la Ca\~nada, Madrid, Spain\label{aff44}
\and
Universit\'e Claude Bernard Lyon 1, CNRS/IN2P3, IP2I Lyon, UMR 5822, Villeurbanne, F-69100, France\label{aff45}
\and
Aix-Marseille Universit\'e, CNRS, CNES, LAM, Marseille, France\label{aff46}
\and
Institut de Ci\`{e}ncies del Cosmos (ICCUB), Universitat de Barcelona (IEEC-UB), Mart\'{i} i Franqu\`{e}s 1, 08028 Barcelona, Spain\label{aff47}
\and
Instituci\'o Catalana de Recerca i Estudis Avan\c{c}ats (ICREA), Passeig de Llu\'{\i}s Companys 23, 08010 Barcelona, Spain\label{aff48}
\and
Institut de Ciencies de l'Espai (IEEC-CSIC), Campus UAB, Carrer de Can Magrans, s/n Cerdanyola del Vall\'es, 08193 Barcelona, Spain\label{aff49}
\and
UCB Lyon 1, CNRS/IN2P3, IUF, IP2I Lyon, 4 rue Enrico Fermi, 69622 Villeurbanne, France\label{aff50}
\and
Mullard Space Science Laboratory, University College London, Holmbury St Mary, Dorking, Surrey RH5 6NT, UK\label{aff51}
\and
Departamento de F\'isica, Faculdade de Ci\^encias, Universidade de Lisboa, Edif\'icio C8, Campo Grande, PT1749-016 Lisboa, Portugal\label{aff52}
\and
Instituto de Astrof\'isica e Ci\^encias do Espa\c{c}o, Faculdade de Ci\^encias, Universidade de Lisboa, Campo Grande, 1749-016 Lisboa, Portugal\label{aff53}
\and
Department of Astronomy, University of Geneva, ch. d'Ecogia 16, 1290 Versoix, Switzerland\label{aff54}
\and
Universit\'e Paris-Saclay, CNRS, Institut d'astrophysique spatiale, 91405, Orsay, France\label{aff55}
\and
INFN-Padova, Via Marzolo 8, 35131 Padova, Italy\label{aff56}
\and
Aix-Marseille Universit\'e, CNRS/IN2P3, CPPM, Marseille, France\label{aff57}
\and
INAF-Istituto di Astrofisica e Planetologia Spaziali, via del Fosso del Cavaliere, 100, 00100 Roma, Italy\label{aff58}
\and
Space Science Data Center, Italian Space Agency, via del Politecnico snc, 00133 Roma, Italy\label{aff59}
\and
INFN-Bologna, Via Irnerio 46, 40126 Bologna, Italy\label{aff60}
\and
School of Physics, HH Wills Physics Laboratory, University of Bristol, Tyndall Avenue, Bristol, BS8 1TL, UK\label{aff61}
\and
Department of Physics, Oxford University, Keble Road, Oxford OX1 3RH, UK\label{aff62}
\and
University Observatory, LMU Faculty of Physics, Scheinerstr.~1, 81679 Munich, Germany\label{aff63}
\and
Max Planck Institute for Extraterrestrial Physics, Giessenbachstr. 1, 85748 Garching, Germany\label{aff64}
\and
Universit\"ats-Sternwarte M\"unchen, Fakult\"at f\"ur Physik, Ludwig-Maximilians-Universit\"at M\"unchen, Scheinerstr.~1, 81679 M\"unchen, Germany\label{aff65}
\and
National Research Council, Herzberg Astronomy and Astrophysics Research Centre, 5071 W. Saanich Rd. Victoria, BC, V9E 2E7, Canada\label{aff66}
\and
Institute of Theoretical Astrophysics, University of Oslo, P.O. Box 1029 Blindern, 0315 Oslo, Norway\label{aff67}
\and
Caltech/IPAC, 1200 E. California Blvd., Pasadena, CA 91125, USA\label{aff68}
\and
Jet Propulsion Laboratory, California Institute of Technology, 4800 Oak Grove Drive, Pasadena, CA, 91109, USA\label{aff69}
\and
Technical University of Denmark, Elektrovej 327, 2800 Kgs. Lyngby, Denmark\label{aff70}
\and
Cosmic Dawn Center (DAWN), Denmark\label{aff71}
\and
 Instituto de Astrof\'{\i}sica de Canarias, E-38205 La Laguna; Universidad de La Laguna, Dpto. Astrof\'\i sica, E-38206 La Laguna, Tenerife, Spain\label{aff72}
\and
Universit\'e PSL, Observatoire de Paris, Sorbonne Universit\'e, CNRS, LERMA, 75014, Paris, France\label{aff73}
\and
Universit\'e Paris-Cit\'e, 5 Rue Thomas Mann, 75013, Paris, France\label{aff74}
\and
Max-Planck-Institut f\"ur Astronomie, K\"onigstuhl 17, 69117 Heidelberg, Germany\label{aff75}
\and
NASA Goddard Space Flight Center, Greenbelt, MD 20771, USA\label{aff76}
\and
Department of Physics and Astronomy, University College London, Gower Street, London WC1E 6BT, UK\label{aff77}
\and
Department of Physics and Helsinki Institute of Physics, Gustaf H\"allstr\"omin katu 2, University of Helsinki, 00014 Helsinki, Finland\label{aff78}
\and
Universit\'e Paris-Saclay, Universit\'e Paris Cit\'e, CEA, CNRS, AIM, 91191, Gif-sur-Yvette, France\label{aff79}
\and
Universit\'e de Gen\`eve, D\'epartement de Physique Th\'eorique and Centre for Astroparticle Physics, 24 quai Ernest-Ansermet, CH-1211 Gen\`eve 4, Switzerland\label{aff80}
\and
Department of Physics, P.O. Box 64, University of Helsinki, 00014 Helsinki, Finland\label{aff81}
\and
Helsinki Institute of Physics, Gustaf H{\"a}llstr{\"o}min katu 2, University of Helsinki, 00014 Helsinki, Finland\label{aff82}
\and
Laboratoire d'etude de l'Univers et des phenomenes eXtremes, Observatoire de Paris, Universit\'e PSL, Sorbonne Universit\'e, CNRS, 92190 Meudon, France\label{aff83}
\and
SKAO, Jodrell Bank, Lower Withington, Macclesfield SK11 9FT, UK\label{aff84}
\and
Centre de Calcul de l'IN2P3/CNRS, 21 avenue Pierre de Coubertin 69627 Villeurbanne Cedex, France\label{aff85}
\and
Dipartimento di Fisica "Aldo Pontremoli", Universit\`a degli Studi di Milano, Via Celoria 16, 20133 Milano, Italy\label{aff86}
\and
INFN-Sezione di Milano, Via Celoria 16, 20133 Milano, Italy\label{aff87}
\and
University of Applied Sciences and Arts of Northwestern Switzerland, School of Computer Science, 5210 Windisch, Switzerland\label{aff88}
\and
Universit\"at Bonn, Argelander-Institut f\"ur Astronomie, Auf dem H\"ugel 71, 53121 Bonn, Germany\label{aff89}
\and
INFN-Sezione di Roma, Piazzale Aldo Moro, 2 - c/o Dipartimento di Fisica, Edificio G. Marconi, 00185 Roma, Italy\label{aff90}
\and
Department of Physics, Institute for Computational Cosmology, Durham University, South Road, Durham, DH1 3LE, UK\label{aff91}
\and
Universit\'e Paris Cit\'e, CNRS, Astroparticule et Cosmologie, 75013 Paris, France\label{aff92}
\and
CNRS-UCB International Research Laboratory, Centre Pierre Bin\'etruy, IRL2007, CPB-IN2P3, Berkeley, USA\label{aff93}
\and
Institute of Physics, Laboratory of Astrophysics, Ecole Polytechnique F\'ed\'erale de Lausanne (EPFL), Observatoire de Sauverny, 1290 Versoix, Switzerland\label{aff94}
\and
Dipartimento di Fisica - Sezione di Astronomia, Universit\`a di Trieste, Via Tiepolo 11, 34131 Trieste, Italy\label{aff95}
\and
Telespazio UK S.L. for European Space Agency (ESA), Camino bajo del Castillo, s/n, Urbanizacion Villafranca del Castillo, Villanueva de la Ca\~nada, 28692 Madrid, Spain\label{aff96}
\and
Institut de F\'{i}sica d'Altes Energies (IFAE), The Barcelona Institute of Science and Technology, Campus UAB, 08193 Bellaterra (Barcelona), Spain\label{aff97}
\and
School of Mathematics and Physics, University of Surrey, Guildford, Surrey, GU2 7XH, UK\label{aff98}
\and
DARK, Niels Bohr Institute, University of Copenhagen, Jagtvej 155, 2200 Copenhagen, Denmark\label{aff99}
\and
Waterloo Centre for Astrophysics, University of Waterloo, Waterloo, Ontario N2L 3G1, Canada\label{aff100}
\and
Department of Physics and Astronomy, University of Waterloo, Waterloo, Ontario N2L 3G1, Canada\label{aff101}
\and
Perimeter Institute for Theoretical Physics, Waterloo, Ontario N2L 2Y5, Canada\label{aff102}
\and
Centre National d'Etudes Spatiales -- Centre spatial de Toulouse, 18 avenue Edouard Belin, 31401 Toulouse Cedex 9, France\label{aff103}
\and
Institute of Space Science, Str. Atomistilor, nr. 409 M\u{a}gurele, Ilfov, 077125, Romania\label{aff104}
\and
Institut f\"ur Theoretische Physik, University of Heidelberg, Philosophenweg 16, 69120 Heidelberg, Germany\label{aff105}
\and
Institut de Recherche en Astrophysique et Plan\'etologie (IRAP), Universit\'e de Toulouse, CNRS, UPS, CNES, 14 Av. Edouard Belin, 31400 Toulouse, France\label{aff106}
\and
Universit\'e St Joseph; Faculty of Sciences, Beirut, Lebanon\label{aff107}
\and
Departamento de F\'isica, FCFM, Universidad de Chile, Blanco Encalada 2008, Santiago, Chile\label{aff108}
\and
Universit\"at Innsbruck, Institut f\"ur Astro- und Teilchenphysik, Technikerstr. 25/8, 6020 Innsbruck, Austria\label{aff109}
\and
Institut d'Estudis Espacials de Catalunya (IEEC),  Edifici RDIT, Campus UPC, 08860 Castelldefels, Barcelona, Spain\label{aff110}
\and
Satlantis, University Science Park, Sede Bld 48940, Leioa-Bilbao, Spain\label{aff111}
\and
Institute of Space Sciences (ICE, CSIC), Campus UAB, Carrer de Can Magrans, s/n, 08193 Barcelona, Spain\label{aff112}
\and
Infrared Processing and Analysis Center, California Institute of Technology, Pasadena, CA 91125, USA\label{aff113}
\and
Instituto de Astrof\'isica e Ci\^encias do Espa\c{c}o, Faculdade de Ci\^encias, Universidade de Lisboa, Tapada da Ajuda, 1349-018 Lisboa, Portugal\label{aff114}
\and
Cosmic Dawn Center (DAWN)\label{aff115}
\and
Niels Bohr Institute, University of Copenhagen, Jagtvej 128, 2200 Copenhagen, Denmark\label{aff116}
\and
Universidad Polit\'ecnica de Cartagena, Departamento de Electr\'onica y Tecnolog\'ia de Computadoras,  Plaza del Hospital 1, 30202 Cartagena, Spain\label{aff117}
\and
Dipartimento di Fisica e Scienze della Terra, Universit\`a degli Studi di Ferrara, Via Giuseppe Saragat 1, 44122 Ferrara, Italy\label{aff118}
\and
Istituto Nazionale di Fisica Nucleare, Sezione di Ferrara, Via Giuseppe Saragat 1, 44122 Ferrara, Italy\label{aff119}
\and
Astronomical Observatory of the Autonomous Region of the Aosta Valley (OAVdA), Loc. Lignan 39, I-11020, Nus (Aosta Valley), Italy\label{aff120}
\and
Universit\'e C\^{o}te d'Azur, Observatoire de la C\^{o}te d'Azur, CNRS, Laboratoire Lagrange, Bd de l'Observatoire, CS 34229, 06304 Nice cedex 4, France\label{aff121}
\and
Dipartimento di Fisica, Sapienza Universit\`a di Roma, Piazzale Aldo Moro 2, 00185 Roma, Italy\label{aff122}
\and
Aurora Technology for European Space Agency (ESA), Camino bajo del Castillo, s/n, Urbanizacion Villafranca del Castillo, Villanueva de la Ca\~nada, 28692 Madrid, Spain\label{aff123}
\and
Zentrum f\"ur Astronomie, Universit\"at Heidelberg, Philosophenweg 12, 69120 Heidelberg, Germany\label{aff124}
\and
ICL, Junia, Universit\'e Catholique de Lille, LITL, 59000 Lille, France\label{aff125}
\and
ICSC - Centro Nazionale di Ricerca in High Performance Computing, Big Data e Quantum Computing, Via Magnanelli 2, Bologna, Italy\label{aff126}
\and
Instituto de F\'isica Te\'orica UAM-CSIC, Campus de Cantoblanco, 28049 Madrid, Spain\label{aff127}
\and
CERCA/ISO, Department of Physics, Case Western Reserve University, 10900 Euclid Avenue, Cleveland, OH 44106, USA\label{aff128}
\and
Technical University of Munich, TUM School of Natural Sciences, Physics Department, James-Franck-Str.~1, 85748 Garching, Germany\label{aff129}
\and
Max-Planck-Institut f\"ur Astrophysik, Karl-Schwarzschild-Str.~1, 85748 Garching, Germany\label{aff130}
\and
Laboratoire Univers et Th\'eorie, Observatoire de Paris, Universit\'e PSL, Universit\'e Paris Cit\'e, CNRS, 92190 Meudon, France\label{aff131}
\and
Departamento de F{\'\i}sica Fundamental. Universidad de Salamanca. Plaza de la Merced s/n. 37008 Salamanca, Spain\label{aff132}
\and
Universit\'e de Strasbourg, CNRS, Observatoire astronomique de Strasbourg, UMR 7550, 67000 Strasbourg, France\label{aff133}
\and
Center for Data-Driven Discovery, Kavli IPMU (WPI), UTIAS, The University of Tokyo, Kashiwa, Chiba 277-8583, Japan\label{aff134}
\and
California Institute of Technology, 1200 E California Blvd, Pasadena, CA 91125, USA\label{aff135}
\and
Department of Physics \& Astronomy, University of California Irvine, Irvine CA 92697, USA\label{aff136}
\and
Departamento F\'isica Aplicada, Universidad Polit\'ecnica de Cartagena, Campus Muralla del Mar, 30202 Cartagena, Murcia, Spain\label{aff137}
\and
Instituto de F\'isica de Cantabria, Edificio Juan Jord\'a, Avenida de los Castros, 39005 Santander, Spain\label{aff138}
\and
INFN, Sezione di Lecce, Via per Arnesano, CP-193, 73100, Lecce, Italy\label{aff139}
\and
Department of Mathematics and Physics E. De Giorgi, University of Salento, Via per Arnesano, CP-I93, 73100, Lecce, Italy\label{aff140}
\and
INAF-Sezione di Lecce, c/o Dipartimento Matematica e Fisica, Via per Arnesano, 73100, Lecce, Italy\label{aff141}
\and
Institute of Cosmology and Gravitation, University of Portsmouth, Portsmouth PO1 3FX, UK\label{aff142}
\and
Department of Computer Science, Aalto University, PO Box 15400, Espoo, FI-00 076, Finland\label{aff143}
\and
Universidad de La Laguna, Dpto. Astrof\'\i sica, E-38206 La Laguna, Tenerife, Spain\label{aff144}
\and
Ruhr University Bochum, Faculty of Physics and Astronomy, Astronomical Institute (AIRUB), German Centre for Cosmological Lensing (GCCL), 44780 Bochum, Germany\label{aff145}
\and
Department of Physics and Astronomy, Vesilinnantie 5, University of Turku, 20014 Turku, Finland\label{aff146}
\and
Serco for European Space Agency (ESA), Camino bajo del Castillo, s/n, Urbanizacion Villafranca del Castillo, Villanueva de la Ca\~nada, 28692 Madrid, Spain\label{aff147}
\and
ARC Centre of Excellence for Dark Matter Particle Physics, Melbourne, Australia\label{aff148}
\and
Centre for Astrophysics \& Supercomputing, Swinburne University of Technology,  Hawthorn, Victoria 3122, Australia\label{aff149}
\and
Department of Physics and Astronomy, University of the Western Cape, Bellville, Cape Town, 7535, South Africa\label{aff150}
\and
DAMTP, Centre for Mathematical Sciences, Wilberforce Road, Cambridge CB3 0WA, UK\label{aff151}
\and
Kavli Institute for Cosmology Cambridge, Madingley Road, Cambridge, CB3 0HA, UK\label{aff152}
\and
Department of Astrophysics, University of Zurich, Winterthurerstrasse 190, 8057 Zurich, Switzerland\label{aff153}
\and
IRFU, CEA, Universit\'e Paris-Saclay 91191 Gif-sur-Yvette Cedex, France\label{aff154}
\and
Oskar Klein Centre for Cosmoparticle Physics, Department of Physics, Stockholm University, Stockholm, SE-106 91, Sweden\label{aff155}
\and
Astrophysics Group, Blackett Laboratory, Imperial College London, London SW7 2AZ, UK\label{aff156}
\and
Univ. Grenoble Alpes, CNRS, Grenoble INP, LPSC-IN2P3, 53, Avenue des Martyrs, 38000, Grenoble, France\label{aff157}
\and
INAF-Osservatorio Astrofisico di Arcetri, Largo E. Fermi 5, 50125, Firenze, Italy\label{aff158}
\and
Centro de Astrof\'{\i}sica da Universidade do Porto, Rua das Estrelas, 4150-762 Porto, Portugal\label{aff159}
\and
HE Space for European Space Agency (ESA), Camino bajo del Castillo, s/n, Urbanizacion Villafranca del Castillo, Villanueva de la Ca\~nada, 28692 Madrid, Spain\label{aff160}
\and
INAF - Osservatorio Astronomico d'Abruzzo, Via Maggini, 64100, Teramo, Italy\label{aff161}
\and
Theoretical astrophysics, Department of Physics and Astronomy, Uppsala University, Box 516, 751 37 Uppsala, Sweden\label{aff162}
\and
Mathematical Institute, University of Leiden, Einsteinweg 55, 2333 CA Leiden, The Netherlands\label{aff163}
\and
School of Physics \& Astronomy, University of Southampton, Highfield Campus, Southampton SO17 1BJ, UK\label{aff164}
\and
ASTRON, the Netherlands Institute for Radio Astronomy, Postbus 2, 7990 AA, Dwingeloo, The Netherlands\label{aff165}
\and
Center for Advanced Interdisciplinary Research, Ss. Cyril and Methodius University in Skopje, Macedonia\label{aff166}
\and
Institute of Astronomy, University of Cambridge, Madingley Road, Cambridge CB3 0HA, UK\label{aff167}
\and
Univ. Lille, CNRS, Centrale Lille, UMR 9189 CRIStAL, 59000 Lille, France\label{aff168}
\and
Department of Astrophysical Sciences, Peyton Hall, Princeton University, Princeton, NJ 08544, USA\label{aff169}
\and
Space physics and astronomy research unit, University of Oulu, Pentti Kaiteran katu 1, FI-90014 Oulu, Finland\label{aff170}
\and
Center for Computational Astrophysics, Flatiron Institute, 162 5th Avenue, 10010, New York, NY, USA\label{aff171}}

\date{Received ; accepted }

  \abstract{We present and investigate the properties of a sample of radio-selected, \Euclid-dark galaxies, identified from LOFAR HBA observations at 144\,MHz within the Euclid Deep Field-North (EDF-N), covering approximately 10 deg$^2$. Starting from \num{1051} radio sources lacking optical counterparts down to a $5\sigma$ depth of $23.9$ in the $y$ band in previous surveys, but detected with \textit{Spitzer}/IRAC, we identified 166 galaxies with no emission at a more than $3\sigma$ level in \Euclid Quick Release 1 (Q1) images, reaching a $5\sigma$ depth of $23.9$ in the $\HE$ band, and no matches in the \Euclid Q1 catalogue.
To minimise contamination from nearby sources, we selected a sub-sample of 88 isolated galaxies.
By exploiting multi-band images and catalogues available for the EDF-N, we inferred the physical properties of our sample via SED-fitting. The resulting redshift distribution spans $0.4 \leq z_\mathrm{ph} \leq 5.0$. We made use of recent sub-arcsecond imaging from the International LOFAR Telescope over an area of $2.5\,\rm deg\times2.5$\,deg, to constrain the nature of the compact radio emission in our targets through brightness temperature estimates. By combining this information with the radio excess relative to the infrared/radio-correlation (IRRC), we searched for possible active galactic nuclei (AGN) activity in our sample. Approximately 30\% of our sources show a radio excess relative to star formation. This fraction increases to 40\% when including AGN identified via brightness temperature.
The \Euclid-dark sources detected in the far-infrared (FIR) are consistent with a population of heavily obscured ($\langle A_V^{\rm ISM} \rangle_{} = 5.0$), massive ($\langle \logten(M_{\star}/M_{\odot}) \rangle_{} = 11.6$) star-forming galaxies (SFGs) with high star formation rates ($\langle \logten({\mathrm{SFR}/M_{\odot}\,\rm yr^{-1})} \rangle_{} = 3.3$). Their location above the star-forming main sequence is consistent with the high-mass and high-SFR end of similar near-infrared (NIR)-dark galaxy populations selected at higher radio frequencies reported in the literature.
We performed a UV-to-radio median stacking analysis, dividing the sources with LOFAR ILT imaging according to the presence of radio excess. Both subsamples are detected in the $\HE$ band, at $3\sigma$ and $9\sigma$ for the radio-excess and non–radio-excess sources, respectively. The two stacked populations exhibit similar global physical properties and differ primarily in their radio emission, with values consistent with those observed for individual sources, albeit with less extreme SFR.
These preliminary results indicate that, compared to similar selections carried out over smaller fields, the wide area covered by \Euclid\ enables the identification of a higher fraction of systems in which intense star formation and AGN activity coexist, likely capturing a key phase of galaxy–black hole co-evolution.
}

   \keywords{Galaxies: active - Galaxies: evolution}

   \maketitle
   \nolinenumbers

\section{Introduction}

Continuum radio emission is an effective tracer of star-formation activity inside galaxies, originating from a combination of the synchrotron emission from electrons accelerated in supernova remnants and free-free continuum radiation from ionised H\,{\sc ii} regions (e.g. \citealt{Condon2002AJ....124..675C}; \citealt{Kennicutt2012AR}). 
Unlike other star-formation tracers, such as the ultraviolet (UV) and H$\,\alpha$ luminosity, the radio emission is unaffected by dust extinction and can be efficiently exploited to study the dust-obscured star-formation (e.g. \citealt{Chapman2004}). Moreover, in contrast to far-infrared (FIR) observations, which often suffer from source-blending issues, radio surveys can achieve sub-arcsecond angular resolution, considerably improving the reliability of counterpart identification.

Radio selection has proven to be extremely effective in identifying populations of dusty star-forming galaxies (SFGs), often missed by optical and near-infrared (NIR) surveys (\citealt{Chapman2001}; \citealt{Smail2002}; \citealt{Talia2021}; \citealt{Enia2022ApJ...927..204E}; \citealt{Behiri2023}; \citealt{vanderVlugt2023ApJ...951..131V}; and \citealt{Gentile2024a, Gentile2025} defined as \citetalias{Gentile2024a} and \citetalias{Gentile2025} hereafter). Because of their faintness at shorter wavelengths, these extreme objects, and more generally all sources with very weak optical/near-infrared (NIR) emission, are commonly named `optically dark' galaxies (\citealt{Simpson2014}; \citealt{WangT2019}; \citealt{Gruppioni2020}; \citealt{Sun2021}; \citealt{Fudamoto2021}; \citealt{Smail2021}; \citealt{Shu2022}), but are also referred to as {\it Hubble} Space Telescope (HST)-dark, radio-selected (Rs)-NIR dark, \textit{H}-dropout, NIR-dark/faint, optical/NIR (OIR)-dark, depending on their selection criteria (\citealt{Wang2016}; \citealt{Franco2018}; \citealt{WilliamsC2019}; \citealt{Gruppioni2020}; \citealt{Riechers2020}; \citealt{Manning2022}). 
Such studies have associated these populations with massive ($M_{\star}>10^{10}\,$M$_{\odot}$), dust-rich ($A_V>3$), and high-redshift ($z>2$--$3$) galaxies, highlighting their potential impact on the cosmic star-formation rate density (SFRD). In particular, their contribution has been estimated to account for up to 40\% of the high-redshift SFRD density traced by Lyman-break galaxies (e.g., \citealt{Talia2021}; \citealt{Barrufet2023}; \citealt{Xiao2023};  \citealt{Xiao2024}; \citetalias{Gentile2025}).

One major caveat to consider when using the radio band as a tracer of star formation is the potential contamination from active galactic nuclei (AGN).
Indeed, in the framework of galaxy and supermassive black hole (SMBH) co-evolution (\citealt{Hopkins2008}; \citealt{Hickox2018}; \citealt{Lapi2018}), star formation and SMBH accretion are likely co-occurring during an obscured phase.
Radio emission provides a complementary probe of AGN activity along with X-ray emission. The radio band can identify dust-obscured sources where X-ray selection becomes increasingly incomplete, particularly in the Compton-thick regime. In addition, radio and X-ray observations probe only partially overlapping AGN populations, making radio data essential for a more complete census of AGN activity (\citealt{Mazzolari2024}). 
In particular, deep radio surveys ($S_{\rm 1.4 \,GHz} \leq 0.1\,$mJy, e.g., \citealt{Mancuso2017}) are also sensitive to radio-quiet (RQ) AGN, which exhibit evidence of nuclear activity in the X-ray, mid-infrared (MIR), or optical bands but show weak radio emission (i.e., no large-scale jets). 
The nature of radio emission within RQ AGN remains a subject of debate, since it likely arises from the interplay of multiple physical mechanisms, including compact jets on (sub-)kiloparsec scales, star-forming regions in the circumnuclear environment, and other AGN-driven processes, such as coronal winds or radiatively driven shocks. This complexity highlights the diverse nature of RQ AGN and suggests that their radio emission cannot be attributed to a single dominant mechanism (see \citealt{Panessa2019} for a review, \citealt{Bonato2017}; \citealt{Mancuso2017}; \citealt{Ceraj2018}; \citealt{Prandoni2018}).

Most of the studies conducted so far on radio-selected optical–NIR dark sources have been based on deep, high-frequency surveys covering relatively small areas such as the Cosmic Evolution Survey (COSMOS; \citealt{Scoville2007}) and the Great Observatories Origins Deep Survey
ﬁeld in the Northern hemisphere (GOODS-N, \citealt{Dickinson2003}; \citealt{Giavalisco2004}). Those include the \textit{Karl G. Jansky} Very Large Array (VLA)-COSMOS 3\,GHz Large Project \citep{Smolcic2017, Talia2021, Behiri2023, Gentile2024a, Gentile2025}, the ultra-deep 3\,GHz VLA observations from the COSMOS-XS survey \citep{vanderVlugt2021ApJ...907....5V, vanderVlugt2022ApJ...941...10V, vanderVlugt2023ApJ...951..131V}, and the 1.4\,GHz VLA observations in the GOODS-N field \citep{Owen2018,Enia2022ApJ...927..204E}.
Building on these previous efforts, our work aims to extend the study of radio-selected NIR dark sources to larger areas by exploiting the unique features of the Euclid Deep Field North (EDF-N). 

Covering approximately $22\,\mathrm{deg}^2$, the EDF-N offers a significant increase in survey area, thus enabling the possibility to detect rare objects as optical-NIR dark galaxies whose number density at $z>3$ is estimated to be approximately $2\times10^{-5}$ Mpc$^{-3}$ (see e.g. \citealt{Schreiber2018b}; \citealt{Valentino2020}; \citealt{Gentile2024a}). Wide-area surveys are therefore crucial to efficiently identify and statistically characterise these rare and extreme sources. In this work, we focus on the $10\,\mathrm{deg}^2$ region covered by Low-Frequency Array (LOFAR) High Band Antenna (HBA) observations at 144\,MHz, along with the available multi-wavelength information, including the \cite{Q1cite} in the optical/NIR regime. This enables the exploration of the radio properties of these sources at much lower frequencies than typically used in past studies. The low-frequency regime is dominated by synchrotron emission, while the thermal free-free becomes negligible. Synchrotron self-absorption might, however, play an important role in suppressing the non-thermal emission (e.g. \citealt{Mancuso2017}; \citealt{Schober2017MNRAS.468..946S}).

The paper is organised as follows. In \cref{sec:data}, we provide an overview of the multi-band data products used in this study. \Cref{sec:selection} describes the selection criteria, while \cref{sec:analysis} details the analysis methods. The main results are presented in \cref{sec:results}, and finally, in \cref{sec:summary}, we discuss the implications of our findings and summarise our conclusions. Throughout the paper, we assume a \cite{Chabrier2003} initial mass function (IMF) and we consider a $\Lambda$CDM cosmology with $H_0=70\,{\rm km}\,{\rm s}^{-1}\,{\rm Mpc}^{-1}$, $\Omega_{\Lambda}=0.7$, and $\Omega_{\rm m}=0.3$,  All magnitudes are reported in the AB system \citep{Oke1983}.

\section{Data description}\label{sec:data}

\subsection{The LOFAR multi-band catalogue}\label{sec:b25cat}

This paper is based on the radio source catalogue presented in \citet[][B25 hereafter]{Bisigello2025b}, which has been produced based on an updated version of the LOFAR HBA radio source catalogue presented by \cite{Bondi2024}, derived from 72 hours of observations in the EDF-N at the frequency $\nu=144$\,MHz. Images reach a spatial resolution of $6\arcsec$ and a $1\sigma$ sensitivity of \SI{32}{\micro\jansky}\,beam$^{-1}$ over 10\,deg$^2$. The catalogue includes radio sources with ${\rm S/N}>5$. The spurious fraction corresponds to around $3\%$ for sources with $5.0 \leq {\rm S/N}<5.5$, around $\sim 1.5\%$ for sources with $5.5\leq {\rm S/N}<6.0$, and less than $0.5\%$ for ${\rm S/N}>10$. \citetalias{Bisigello2025b} identified optical and/or MIR counterparts\footnote{In \citetalias{Bisigello2025b}, the regime covered by \textit{Spitzer} bands is defined as NIR. In this paper, we refer to this wavelength range as MIR instead.} by exploiting the optical catalogues from the Hawaii eROSITA Ecliptic Pole (HEROES, \citealt{Taylor2023}) survey which reach $5\sigma$ depths of $g=26.5$, $r=26.2$, $i=25.7$, $z=25.1$, $y=23.9$, $\rm NB816=24.4$, and $\rm NB921=24.4$ and from MIR \textit{Spitzer} maps from the Cosmic Dawn Survey \citep{Moneti-EP17, EP-McPartland}. 
\citetalias{Bisigello2025b} produced a new \textit{Spitzer}/Infrared Array Camera (IRAC) source catalogue by performing a new source extraction aimed at maximising deblending.  Moreover, compared to the catalogue by \cite{Moneti-EP17}, which included only the two shorter-wavelength IRAC bands, the \citetalias{Bisigello2025b} catalogue combines the available observations in all four IRAC channels (3.6, 4.5, 5.8, and 8.0$\,\micron$), offering a more complete view of redder and potentially more obscured sources. Here we provide a summary of the extraction procedure, and we refer to \citetalias{Bisigello2025b} for the detailed description.
In brief, \citetalias{Bisigello2025b} exploited the \texttt{PHOTUTILS} Python package \citep{Bradley2023} to create a segmentation map from the co-added $3.6\,\micron$ and $4.5\,\micron$ images, taking into account the respective error maps. They selected as sources all regions containing at least three connected pixels above 3 times the local RMS noise. A deblending procedure was then applied, where peaks were identified as separate sources if they contained at least two pixels each above the flux threshold and exceeded a minimum contrast, defined as the fraction of the total source flux in the local peak. The contrast parameter ranges from 0 (all local peaks treated as separate sources) to 1 (no deblending); a very low contrast of $10^{-8}$ was chosen to maximise the deblending capability.\footnote{We verified that the number of deblended sources remains unchanged for contrast values between $10^{-4}$ and $10^{-8}$, and visual inspection confirmed that no over-segmentation is introduced.} This setup was fine-tuned by visually inspecting a subsample of potentially blended sources, confirming that individual extended sources were not erroneously split into multiple components.
 Finally, the segmentation map was used as a reference to each band separately to extract fluxes via aperture photometry. This approach allowed for the inclusion of objects detected only in the IRAC bands, which is particularly relevant for identifying redder, potentially more obscured counterparts to radio sources, as shown by previous studies \citep[e.g.,][]{Williams2019}. 

Additionally, \citetalias{Bisigello2025b} included the \textit{u} band observations that are also part of the Cosmic Dawn Survey \citep{EP-McPartland}.
Counterpart identification was carried out using a likelihood ratio method (\citealt{deRuiter1977}; \citealt{Sutherland1992}), complemented by visual inspection. This combined approach enabled the reliable identification of both isolated and compact radio sources, as well as those with complex or extended morphologies, including those in crowded fields. \citetalias{Bisigello2025b} also exploited LOFAR intermediate-resolution data (1\arcsecf5) obtained with the International LOFAR Telescope (ILT), including the international LOFAR stations. These observations cover an inner region of the EDF-N of $2.5\,\rm deg\times2.5\,\rm deg$ and combine 32 out of the 72\,h available, reaching a central RMS noise of \SI{36}{\micro\jansky}\,beam$^{-1}$ and were used to improve the accuracy of previous radio positions. As a result of their analysis, 99.2$\%$ of the radio sources were associated with a reliable optical and/or MIR counterpart, corresponding to \num{19401} objects out of a total of \num{19550}. Moreover, the \citetalias{Bisigello2025b} catalogue also includes FIR data obtained by cross-matching the LOFAR sources with photometry from the \textit{Herschel} Space Observatory, retrieved from the IPAC infrared science archive. Specifically, LOFAR sources were searched for counterparts in the Photodetector Array Camera and Spectrometer (PACS) Point Source Catalogue \citep{Marton2017} at 70, 100, and 160\,$\micron$, and in the Spectral and Photometric Imaging Receiver (SPIRE) Point Source Catalogue \citep{Schulz2018} at 250, 350, and 500\,$\micron$. A matching radius of $6\arcsec$ was adopted for PACS, and $12\arcsec$ for SPIRE. As a result of this cross-match, approximately 20\% of the LOFAR sources have FIR detections in at least one of the \textit{Herschel}/SPIRE bands. In \cref{sec:phot_isolated} we describe the analysis of FIR data and the cross-match performed specifically for our sample.

In the following sections, we describe the additional datasets used in this paper, whose properties are summarised in \cref{tab:filters}.

\subsection{\Euclid optical and NIR data}

The \Euclid mission (see \citealt{EuclidSkyOverview} for an overview) will provide high-resolution optical imaging, as well as NIR imaging and spectroscopy, over about \num{14000} deg$^2$ of extragalactic sky. The first set of data was released on March 19, 2025 (Quick Release 1 or Q1).
A detailed description of the Q1 data release is provided by \citet{Q1-TP001}. Data products and processing from the Visible Camera (VIS) and Near-Infrared Spectrometer and Photometer (NISP) are reported in \citet{Q1-TP002} and \citet{Q1-TP003}, respectively. The VIS instrument observes in the visible \citep[$\IE$,][]{EuclidSkyVIS}, while NISP includes three NIR filters \citep[$\YE$, $\JE$, and $\HE$ band, see][]{EuclidSkyNISP}. The photometric catalogue is presented in \citet{Q1-TP004}. 
The EDF-N is covered by Q1 observations over $22.9\, \mathrm{deg}^2$ in all four photometric bands.
Ground-based data products have also been included along with \Euclid bands. In this work, we exploited images from the Ultraviolet Near-Infrared Optical Northern Survey \citep[UNIONS,][]{Gwyn2025arXiv250313783G}, in particular observations from the Subaru Telescope Hyper Suprime Cam (HSC) and the Canada-France-Hawaii Telescope (CFHT), ranging between $0.3\,\micron$ and $0.9\,\micron$.
The complete list of filters available in each field, with their effective wavelengths, their corresponding observational depths, full width at half maxima (FWHM), and area coverage is reported in \cref{tab:filters}.

\begin{table}[]
    \centering
    \caption{Filters used in this work, with their associated effective wavelength, observed depths, FWHM, and area of the EDF-N covered.}\label{tab:filters}
    \resizebox{0.49\textwidth}{!}{
    \begin{tabular}{lcccc}
    \hline \hline
        \noalign{\vskip 2pt}
        Band &  $\lambda_{\rm eff}$  &  Depth & FWHM & Area\\
         & [\micron] & &[arcsec] & [deg$^2$]\\
        \hline
        \noalign{\vskip 2pt}
        CFHT/MegaCam $u$      & $0.372$ & $24.2$\,mag & $0.89$ & 13.7\\
        HSC $g$               & $0.480$ & $25.7$\,mag & $0.90$& 5.4\\
        CFHT/MegaCam $r$      & $0.640$ & $24.8$\,mag & $0.71$& 13.7\\
        \Euclid VIS/\IE               & $0.715$ & $25.45$\,mag & $0.16$  & 22.9\\
        HSC $z$               & $0.891$ & $24.1$\,mag & $0.71$  & 5.4\\
        \Euclid NISP/\YE              & $1080.9$ & $24.6$\,mag &$ 0.3$ & 22.9\\ 
        \Euclid NISP/\JE              & $1367.3$ & $24.1$\,mag & $0.3$  & 22.9\\
        \Euclid NISP/\HE              & $1771.4$ & $23.9$\,mag & $0.3$  & 22.9\\
        \textit{Spitzer}/IRAC1 	      & $3.550$ & $24.8$\,mag & $1.66$--$1.95^a$ & 11.74\\
       \textit{Spitzer}/IRAC2 	      & $4.493$ & $24.8$\,mag & $1.72$--$2.02^a$ & 11.54\\
        \textit{Spitzer}/IRAC3        & $5.696$ & $20.8$\,mag & $1.88$ & 0.61\\
        \textit{Spitzer}/IRAC4        & $7.799$ & $21.9$\,mag & $1.98$& 0.62\\
        WISE $W3$                 &  $12.082$ & $19.1$\,mag & $8.5$ & All sky\\
        WISE $W4$                 &  $22.194$ & $17.2$\,mag & $17$ & All sky\\
        \textit{Spitzer}/MIPS24 & $23.7$ & $16.95$\,mag & 6 & $0.4$ \\
        AKARI/FIS65           & $65$ & $3.2$\,Jy& $37$ & All sky \\
        AKARI/FIS90           & $90$ &  $0.55$\,Jy & $39$& All sky\\
        AKARI/FIS140          & $140$ & $3.8$\,Jy & $58$ & All sky\\
        AKARI/FIS160          & $160$ & $7.5$\,Jy & $61$ & All sky\\
        \textit{Herschel}/SPIRE250 & $250$ & $4.5\times10^{-2}$\,Jy & $18.2$ &16.05\\
        \textit{Herschel}/SPIRE350 & $350$ & $3.8\times10^{-2}$\,Jy & $24.9$ &16.05\\
        \textit{Herschel}/SPIRE500 & $500$ & $5.4\times10^{-2}$\,Jy & $36.3$ & 16.05\\
        JCMT/SCUBA-2 & $850$ & $(5$--$11.5)\times10^{-3}$\,Jy &$14.6$ &$2.0$\\
    \hline
    \end{tabular}}
    \tablefoot{Reported magnitudes are the $5\sigma$ observed depths. Optical and \Euclid magnitudes have been computed adopting a $2\times\rm FWHM$ diameter aperture and correspond to the median depths of the observing tiles \citep{Q1-TP004}. IRAC1 and IRAC2 depths refer to average values in the fields derived considering $2\arcsec$ empty apertures (see \citealt{Moneti-EP17} and \citealt{EP-McPartland}). IRAC3 and IRAC4 depths are derived from \citetalias{Bisigello2025b} by averaging depths derived from the catalogue, after correcting from aperture to total magnitudes. AKARI depths refer to the 5$\sigma$ noise level per scan in the normal mode (see \citealt{Kawada2007}). $^a$ Numbers refer to the cryogenic and warm missions, respectively. }
\end{table}

\subsection{Mid-IR data}\label{sec:MIR_data}

Along with the \textit{Spitzer}/IRAC photometry described in \cref{sec:b25cat}, we exploited \textit{Spitzer}/MIPS mosaics available in the North Ecliptic Pole (NEP), covering a limited area of 0.4 deg$^2$ \citep{Jarrett2011ApJ...735..112J} and part of the Spitzer Enhanced Imaging Products (SEIP). 
The Wide-field Infrared Survey Explorer (WISE; \citealt{Wright2010}) provides additional MIR coverage over the full sky. In this work, we also exploited data from the AllWISE and unWISE programmes. AllWISE combines data from the WISE cryogenic and NEOWISE \citep{Mainzer2011ApJ...743..156M} post-cryogenic missions and includes both a source catalogue and an image atlas in four photometric bands, $W1$, $W2$, $W3$, and $W4$, centred at 3.4, 4.6, 12, and 22\,$\micron$. The unWISE products \citep{Lang2014AJ....147..108L} collect coadded WISE images that have not been blurred and retain the intrinsic resolution of the data.
Since the first two bands overlap with and are shallower than IRAC1 and IRAC2, we focused on WISE $W3$ and $W4$ (12$\, \micron$ and 22\,$\micron$) to complement the IRAC photometry.

\subsection{Far-IR data}

We made use of multiple FIR and sub-millimetre (mm) data products in the North Ecliptic Pole (NEP) field. First, we included the AKARI/Far Infrared Surveyor (FIS) All-Sky Survey Bright Source Catalogue (\citealt{Yamamura2009AIPC.1158..169Y}), derived from the AKARI Far-infrared All-Sky Survey (\citealt{Doi2015PASJ...67...50D}), which provides observations at 65, 90, 140, and 160\,$\micron$ over the NEP. In addition to the PACS and SPIRE point-source catalogues used in \citetalias{Bisigello2025b}, we utilised the products available from the \textit{Herschel} Extragalactic Legacy Project (HELP; \citealt{Shirley2019, Shirley2021}), including SPIRE maps and blind catalogues from the \textit{Herschel}-AKARI NEP Deep Survey, covering 9.2 deg$^2$ (\citealt{Pearson2017a}). Finally, we made use of data from the James Clerk Maxwell Telescope (JCMT) SCUBA-2, in particular from the North Ecliptic Pole SCUBA-2 (NEPSC2) 850\,$\micron$ survey \citep{Shim2020MNRAS.498.5065S}, covering over 2 deg$^2$ of the NEP.

\subsection{LOFAR ILT 0\arcsecf3 data}

We then exploited recent ILT LOFAR imaging at sub-arcsecond resolution (0\arcsecf3) to investigate the nature of the radio emission of our sources. The map covers the inner $2.5\,\rm deg\times2.5\,\rm deg$ of the EDF-N, reaching a central sensitivity of \SI{32}{\micro\jansky}\,beam$^{-1}$. The wide-field image and the radio catalogue associated with these data will be presented in a future paper (Bondi et al., in prep.). At 144~\,MHz, the sub-arcsec resolution achieved by LOFAR ILT is sensitive to compact radio emission associated with AGN (see e.g. \citealt{Morabito2022, Morabito2025b}) and, as discussed throughout this work, provides a key diagnostic to identify AGN activity in our sources.

\section{Sample selection}\label{sec:selection}

Our selection started from a subsample of \citetalias{Bisigello2025b}’s catalogue, denominated `IRAC-only' sources. This sample comprises \num{1363} objects, corresponding to approximately $7\%$ of the total number of LOFAR sources, and results from \citetalias{Bisigello2025b}’s cross-matching procedure. It is made of sources that lack a reliable counterpart in the optical bands of the HEROES catalogue and have a counterpart in the IRAC bands instead.
Among these sources, although undetected in the optical bands of the HEROES catalogue ($g, r, i, z, y$), \citetalias{Bisigello2025b} identified \num{218} with a counterpart in the $u$ band of the Cosmic Dawn Catalogue. They suggested that this could indicate that some IRAC-only sources lie near the sensitivity limit of the optical data, where more advanced extraction techniques might recover faint optical fluxes. Furthermore, \citetalias{Bisigello2025b} found 94 spurious associations via visual inspection, largely caused by IRAC blending or by the over-segmentation of extended nearby galaxies.
This latter issue arises from the optical catalogue’s source extraction setup, which favours compact sources and may erroneously split a single large galaxy into multiple components, leading to incorrect cross-matches.
Thus, we remove the above-mentioned objects from our starting sample of IRAC-only sources, obtaining a total of \num{1051} IRAC-only radio sources.

\subsection{Cross-match and visual inspection of \Euclid Q1 products}

In addition to the visual inspection conducted in \citetalias{Bisigello2025b}, we analysed the remaining \num{1051} IRAC-only radio sources (cleaned IRAC-only sample) exploiting \Euclid Q1 data products.
We proceeded as follows: 
\begin{itemize}
    \item We overlapped \Euclid, LOFAR, and IRAC positions and the respective 3 and 5$\sigma$ contours to \Euclid VIS and NISP maps. We then excluded all the sources whose LOFAR 1\arcsecf5 and/or IRAC2 emission at 5$\sigma$ contained a $5\sigma$ NIR source in correspondence of the radio and MIR centroid, if present, in at least one of the \Euclid bands. In other words, following a similar approach to that applied by \cite{Talia2021} and \cite{Behiri2023}, our sample only includes sources without a clear $5\sigma$ \Euclid detection corresponding to the radio/IRAC source. As a result of this initial step, a total of \num{304} candidate dark sources were identified.
    \item Secondly, we cross-matched both IRAC and radio positions of this sample with sources in the \Euclid catalogue. We selected a matching radius of 0\arcsecf9, corresponding to around $3\times$ the average positional error of LOFAR positions, which is the instrument with the lowest angular resolution. 
    We chose to match both LOFAR and IRAC positions to \Euclid independently, given that in some cases the separation between LOFAR low resolution and IRAC coordinates can be large (e.g. above $1\arcsec$ for 50/\num{304} sources in our sample), and multiple NIR counterparts might be found within the IRAC or LOFAR beams. 
    As a result of this cross-match, we obtained \num{164} objects without and \num{140} with a counterpart in the \Euclid catalogue. Among the sources with a \Euclid counterpart, nine are flagged as spurious in the catalogue (\texttt{SPURIOUS\_FLAG}=1). We did not include these objects in our sample, since their association with radio and/or MIR emission suggests that they may correspond to real sources rather than artefacts.
    \item We then performed a second visual inspection. In this step, we carefully checked the remaining \num{131} sources with a \Euclid counterpart to verify that the NIR coordinate was correctly associated with the same radio/MIR position. At the same time, for the sources without a \Euclid counterpart, we inspected the images to ensure that there was no significant emission (above $3\sigma$) in the vicinity of the IRAC/radio centroid.
    We found only two cases in which the radio and/or MIR positions are associated with different NIR counterparts, corresponding to objects whose NIR position was wrongly cross-matched with the IRAC2 or LOFAR (low resolution) one because of blending. These sources can therefore be considered \Euclid `dark' (E-dark hereafter) and were added to the sample of objects without a counterpart in the \Euclid catalogue.

\end{itemize}

To summarise, with the above selection, we found a sample of \num{166} E-dark LOFAR-selected targets. We found no evidence for strong gravitational lensing among our sources.

\begin{figure*}
    \centering
    \includegraphics[width=0.8\linewidth]{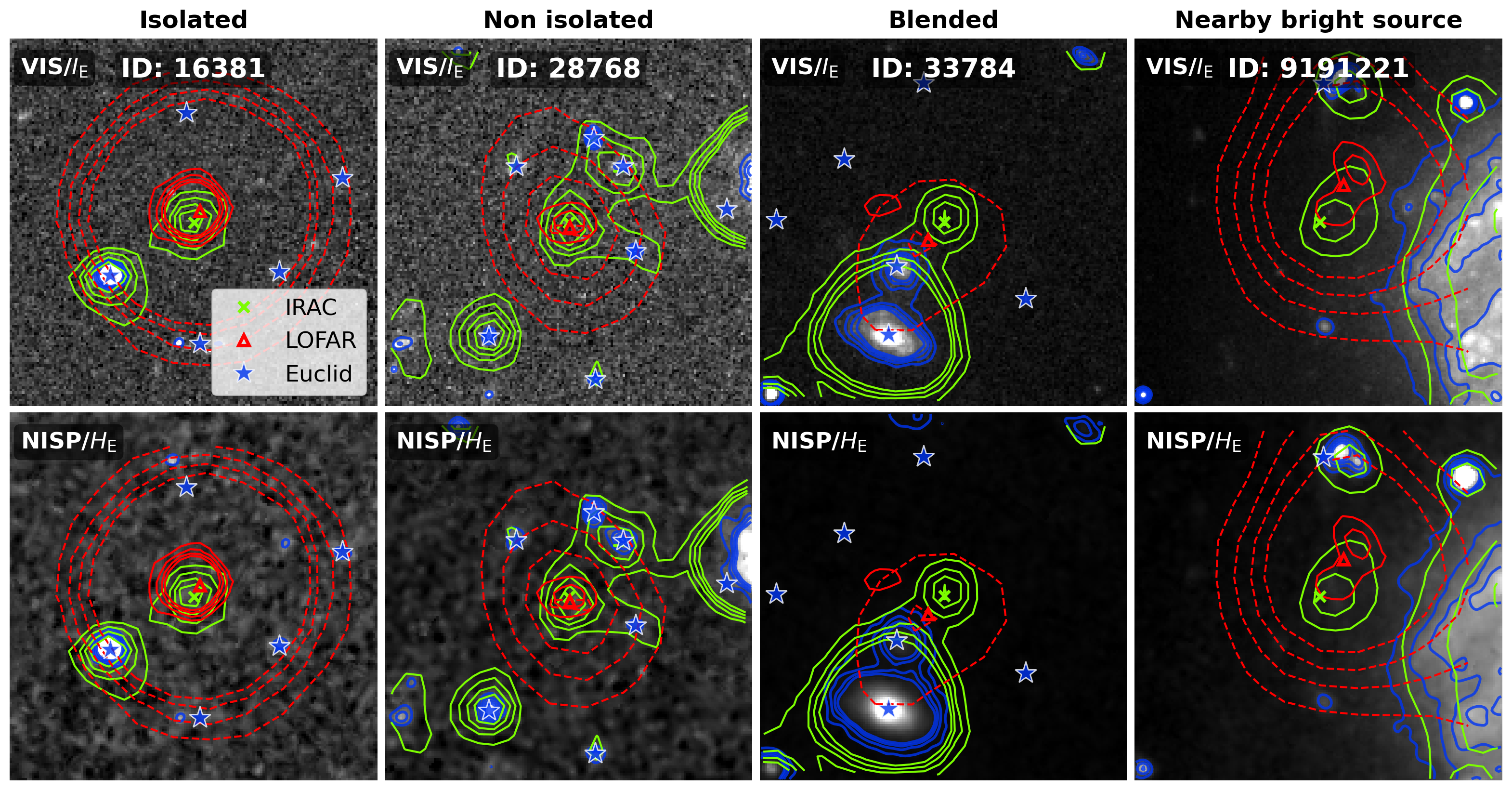}
    \caption{Examples of four E-dark sources illustrating the classification used in this work: isolated sources (satisfying the criteria described in \cref{sec:selection_isolated}), non-isolated sources, blended sources, and sources affected by a nearby bright object.
     Cutouts display the VIS \IE (top panels) and NISP \HE (bottom panels) images centred at the MIR position of the sources in our sample. The {blue, green, and red contours represent the 3, 5, 7, and 9$\sigma$ emission from \Euclid, IRAC2, and LOFAR, respectively, where for LOFAR the intermediate-resolution emission is shown with solid contours and the low-resolution emission with dashed contours.} The centroid of the emission on each instrument is shown, with the corresponding colour, with stars, crosses, and triangles for \Euclid, IRAC2, and LOFAR. The ID of the object matches the one listed in the \citetalias{Bisigello2025b} catalogue. Postage stamps are $15\arcsec\times 15\arcsec$.}
    \label{fig:isolated_source}
\end{figure*}
\subsection{Selection of the `isolated' sample}\label{sec:selection_isolated}

Starting from the parent sample of \num{166} E-dark radio-selected (RS) targets described in the previous section, we define here a sub-sample, hereafter referred to as the `isolated' sample.
This subset is specifically tailored to our preliminary analysis, as it minimises the contamination from nearby objects, at least up to IRAC2, allowing for a cleaner interpretation of their properties. The isolated sample consists of all the E-dark targets whose MIR/radio emission contours at $3\sigma$ from IRAC2, or the LOFAR intermediate-resolution map when available, do not overlap with the $3\sigma$ NIR/MIR emission of nearby IRAC2 or \Euclid sources. This criterion removes about 30\% of the sources inside the sample of 166 objects. Moreover, we excluded sources with an evident MIR or radio blending (corresponding to approximately 11\% of the E-dark sample), located near an extended bright local source or a spike originating from a nearby star (around 5\% of the E-dark sample). From visual inspection, we collected a sample of \num{88} isolated E-dark sources. In \cref{fig:isolated_source} we show examples of four such E-dark sources according to our classification criteria, including a representative isolated source that meets the conditions described above. In \cref{tab:sample_selection} we summarise the main steps described in this section.

The sources in the isolated sample span a flux density range of $0.15 \rm \,mJy \leq S_{\rm 144\,MHz} \leq 50\,mJy$. However, the vast majority of them (about 93\%) have $S_{\rm 144\,MHz} \leq2$\,mJy. This flux regime falls into the region of the radio differential number counts where SFGs and RQ AGN begin to dominate over the radio-loud AGN population \citep[e.g.,][]{Mancuso2017}.

\begin{table}
\centering
\caption{Summary of sample selection steps described in \cref{sec:selection}}
\begin{tabular}{l S[table-format=4]}
\hline
Selection step & {Number of sources} \\
\hline
IRAC-only cleaned (\citetalias{Bisigello2025b}) & 1051 \\
Visual inspection & 304 \\
\Euclid catalogue cross-match & 164 \\
Visual inspection, \Euclid mismatch & 2 \\
Total E-Dark & 166 \\
Isolated E-Dark  & 88 \\
\hline
\end{tabular}
\label{tab:sample_selection}
\end{table}

\section{Analysis}\label{sec:analysis}

\subsection{Photometric counterparts of isolated E-dark sources}\label{sec:phot_isolated}

Besides the photometry available from the \citetalias{Bisigello2025b} catalogue, we made use of data described in \cref{sec:data} to enrich the available photometric information of the sample of isolated E-dark sources.
For optical/NIR bands (i.e., $u$, $g$, $r$, $z$, \IE, \YE, \JE, and \HE) no source is detected by definition and we derived a $3\sigma$ upper limit by measuring the flux inside approximately $100$ random apertures in a $30\arcsec\times30\arcsec$ cutout, avoiding bright sources, and then computing their standard deviation.
In the MIR, besides the IRAC1 and IRAC2 detections, five (three) objects have counterparts in IRAC3 (IRAC4), while 16 sources are covered by MIPS observations, showing no counterpart. For these latter sources, we computed $3\sigma$ upper limits as described above.

We manually extracted WISE fluxes at IRAC positions by adopting \texttt{PHOTUTILS} and by following the procedure reported in the Explanatory Supplement to the AllWISE Data Release Products.\footnote{\url{https://wise2.ipac.caltech.edu/docs/release/allwise/expsup/index.html}} We performed aperture photometry using the fixed circular apertures and relative correction factors reported in section 4.2 of the Explanatory Supplement to the WISE
All-Sky Data Release Products.\footnote{\url{https://wise2.ipac.caltech.edu/docs/release/allsky/expsup/index.html}} For background estimation, we used a standard annulus with inner and outer radii of $50\arcsec$ and $70\arcsec$, respectively, similarly to the values adopted in the AllWISE source catalogue, and masking pixels above the $3\sigma$ noise level. We found respectively two and three detections for WISE3 and WISE 4. Upper limits for the remaining targets were computed in a $400\arcsec\times400\arcsec$ cutout.
For AKARI, there are no individual detections from AKARI/FIS All-Sky Survey Bright Source Catalogue, therefore, we computed the upper limits from the $3\sigma$ rms of the individual $900\arcsec\times900\arcsec$ cutouts centred at the source's position.

For SPIRE, we followed the approach described in \cite{Wang2021}. We first performed a cross-match between our isolated E-dark sources and the HELP blind catalogues (\citealt{Shirley2021}). Initially, we adopted a search radius of $6\arcsec$. For sources with a match within this radius, we then repeated the cross-match using a radius of $18\arcsec$, corresponding to the FWHM of the SPIRE $250\,\micron$ beam. If the matches were unique, the fluxes from the blind photometry catalogues were adopted for further analysis. For sources with multiple possible counterparts within the SPIRE beam, we assigned the SPIRE fluxes using the probabilistic deblending tool \texttt{XID+} (\citealt{Hurley2017}; \citealt{Pearson2017, Pearson2018A&A...615A.146P}), applied to regions around the LOFAR sources on the SPIRE maps. We refer to \cite{Wang2021} for further details on this procedure. Combining this photometry with the SPIRE point-source catalogue, we obtained \num{17}, \num{20}, and \num{16} detections at 250, 350, and 500 \micron, respectively.
We also note that, from the cross-match performed in \citetalias{Bisigello2025b}, no PACS counterparts are available for our sample.
Finally, we cross-matched the 88 isolated E-dark objects with the NEPSC2 catalogue, adopting a search radius of 15\arcsec, which corresponds to the effective beam FWHM of SCUBA-2 at 850\,$\micron$, finding two counterparts.

\subsection{Compact radio emission in the ILT-LOFAR subsample}

We searched for counterparts in the LOFAR ILT map at 0\arcsecf3 resolution. Among the \num{88} sources, \num{50} are covered by the central $1.25\rm \, deg\times 1.25\, deg$ region, reaching the highest sensitivity. We searched for counterparts within 1\arcsec from the IRAC position, finding \num{28} sources with $\rm S/N>4$. The remaining \num{22} objects are not detected at 0\arcsecf3. Moreover, we also searched for additional counterparts in the least sensitive regions of the 0\arcsecf3 radio map, where the rms noise reaches \SI{64}{\micro\jansky}\,beam$^{-1}$. We found three sources in this area, all with $\rm S/N> 5$. Source sizes were obtained by fitting a single Gaussian component to the brightness distribution at 0\arcsecf3 resolution of the detected E-dark galaxies. Most of the detections are consistent with a point-like morphology, with a few exceptions (i.e., ID 17587, 12515, 9319351, 27761), which show a multi-component morphology in the high- and intermediate-resolution LOFAR maps.

\begin{figure}
     \centering
    \includegraphics[width=0.48\textwidth]{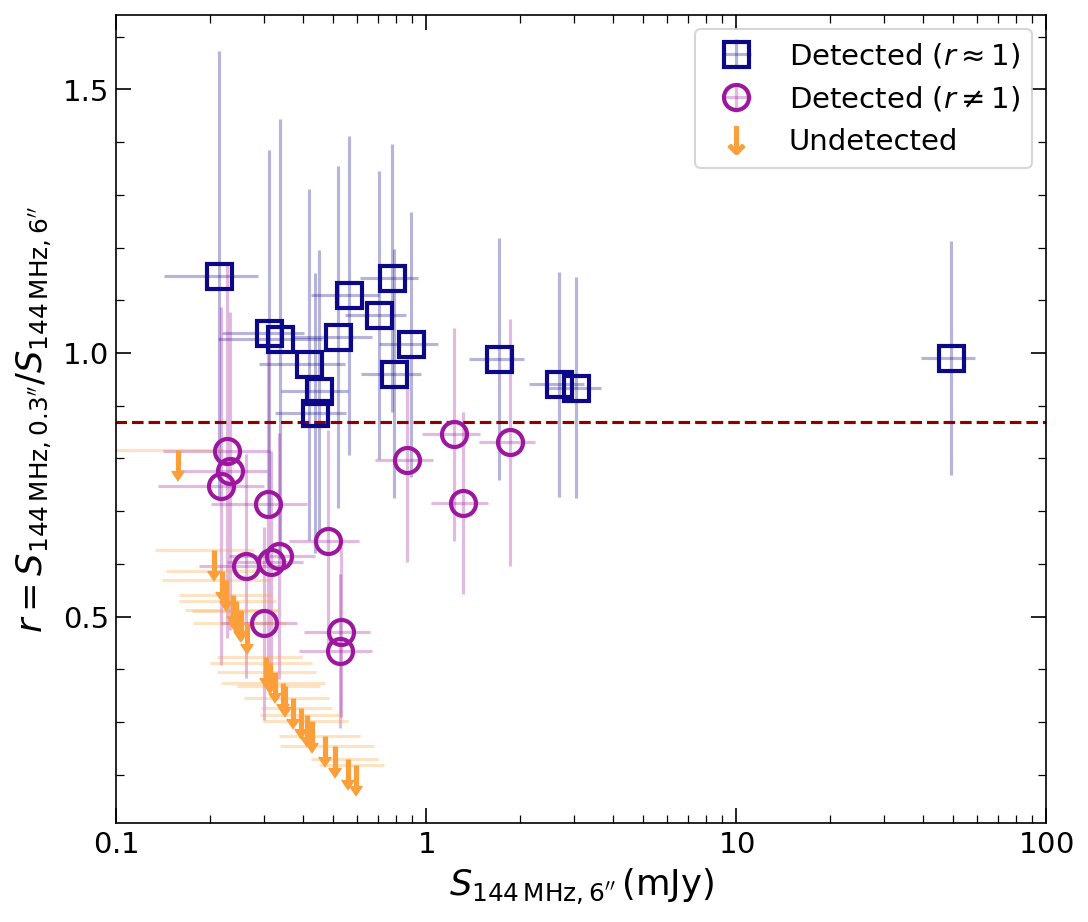}
    \caption{Ratio between 144\,MHz LOFAR fluxes at high (0\arcsecf3) and low (6\arcsec) resolution vs. the 144\,MHz LOFAR flux at 6\arcsec. The dashed horizontal red line is shown as a reference for the aperture flux ratio consistent with unity (represented with blue squares) within the flux uncertainties, corresponding to approximately $15\%$ of the LOFAR flux at 6\arcsec. Sources detected in the LOFAR ILT image are highlighted in blue and purple, corresponding to flux ratios consistent and not consistent with unity, respectively. Yellow arrows mark the $3\sigma$ upper limits for the \num{22} sources without a detection in the LOFAR ILT map.}
    \label{fig:radioflux_ratios}   
\end{figure}

A first indication of the nature of the radio emission can be obtained by comparing the high- and low-resolution LOFAR flux densities. In \cref{fig:radioflux_ratios}, we show the ratio $r$ between the 0\arcsecf3 and 6\arcsec\, radio fluxes as a function of the 6\arcsec\, flux density. Out of the \num{53} sources covered by the LOFAR ILT high-resolution image, \num{16} display flux ratios consistent with unity within the uncertainty on the radio flux at 6\arcsec (corresponding to ratios greater than 0.87). This suggests a dominant radio component and, therefore, that a significant fraction of the radio emission is concentrated on sub-arcsecond scales. For \num{15} objects, the ratio is lower than 0.87, indicating that part of the extended emission is lost at higher angular scales. In these cases, the radio emission could originate from both AGN and star formation processes. For the remaining \num{22} sources without a detection at 0\arcsecf3, we show $3\sigma$ upper limits, suggesting that their radio emission is likely dominated by star formation rather than by an AGN. While this information alone is not sufficient to firmly establish or exclude the presence of AGN activity, it motivates a more quantitative analysis of the compact radio emission and provides useful constraints for the radio component of the spectral energy distribution (SED) fitting for these sources, as discussed in the following section.

As introduced in the previous section, LOFAR ILT 0\arcsecf3 data can be exploited to search for compact emission associated with AGN activity by measuring the radio brightness temperature $T_{\rm b}$. Radio emission in SFGs is expected to have an upper limit in surface brightness \citep{Condon2002AJ....124..675C}, meaning that values surpassing this threshold can only be produced by AGN-related processes. The brightness temperature scales as $T_{\rm b}\propto \nu^{-2}\theta^{-2}$, where $\nu$ is the observing frequency and $\theta$ the angular resolution. At low frequencies, such as 144~MHz, sub-arcsecond resolution observations (0\arcsecf3, corresponding to approximately 2 kpc at a reference redshift of $z=3$) have proven to be sufficient to provide reliable $T_{\rm b}$ measurements \citep{Morabito2022}. Moreover, this approach can be exploited to probe AGN activity even in sources that do not show a clear radio excess relative to the SFR \citep{Morabito2025}. Since the estimate of $T_{\rm b}$ requires knowledge of the source redshift, we combined the radio measurements with SED fitting to derive the physical properties of the sample.

\vspace{0.5cm}

\subsection{SED-fitting}\label{sec:sed_fitting}

We modelled the SED of our sample using the Code Investigating GAlaxy Emission (\texttt{CIGALE}, \citealt{Boquien2019}). \texttt{CIGALE} is a Python SED fitting code able to reproduce broadband UV-to-radio photometric data according to the energy balance (i.e., the energy absorbed from the stellar UV-to-NIR emission is the same as the one re-emitted by the dust in the MIR and FIR regime).
\texttt{CIGALE} estimates the physical properties by comparing modelled galaxy SEDs, generated from a user-defined grid of input parameters, to observed ones. For each parameter, a probability distribution function (PDF) is constructed based on a $\chi^2$ comparison of the observed SED with all models in the n-dimensional grid. The output values include the ‘best’ model, corresponding to the minimum $\chi^2$ point, and the Bayesian-like estimates, obtained as the likelihood-weighted mean of the PDF. The associated uncertainties are computed as the likelihood-weighted standard deviation.
The stellar emission was computed following the \citet[][]{Bruzual2003} population synthesis models, associated with a \cite{Chabrier2003} IMF and metallicity value of $Z=0.02$. We tested additional values for the metallicity (i.e., $Z=0.008$ and $Z=0.05$), finding no significant differences in the derived physical properties.
We also included the nebular emission model, based on \cite{Inoue2011} templates, using the standard values provided by \texttt{CIGALE}.
We assumed a delayed star-formation history (SFH) with a main star-formation episode and an optional burst/quench, quantified by the parameter $t_{\rm bq}$ (i.e., the age of the burst/quench episode). The SFR evolves as
\begin{equation}
    \mathrm{SFR}(t) \propto \frac{t}{\tau^2} \, \exp{\left(-\frac{t}{\tau}\right)}\,, 
\end{equation}
where $t$ is the cosmic time and $\tau$ is the time at which the SFR peaks. 

We modelled the dust attenuation in the far-UV–optical regime using the modified approach of \citet{Charlot2000}, which introduces an age-dependent attenuation described by two separate power laws: one for the diffuse interstellar medium (ISM) and one for the stellar birth clouds (BC). Both components assume a fixed attenuation slope of $\diff A_{\lambda}/\diff \lambda = -0.7$, and the \textit{V}-band attenuation is computed as
\begin{equation}
    \mu = \frac{A_{V}^{\rm ISM}}{A_{V}^{\rm ISM} + A_{V}^{\rm BC}}\; .
\end{equation}
We assumed $A_{V}^{\rm ISM}\in[0.5, 6.0]$ and the default value of $\mu = 0.44$. 
Dust emission is modelled using the \cite{Dale2014} templates, parametrised by the $\alpha_{\rm d}$ parameter defined as $\diff M_{\rm d}(U) \propto U^{-\alpha_{\rm d}}\diff U$, where $M_{\rm d}$ is the dust mass and $U$ the intensity of the radiation field.
The AGN contribution to the total galaxy luminosity (UV-to-radio) is taken into account assuming the templates from \cite{Fritz2006}. By specifying a wavelength range in the parameter $\lambda_{\rm fracAGN}$, it is possible to account for all the components in the computation of the AGN fraction \citep{Yang2022ApJ...927..192Y}, which is defined as
\begin{equation}
    f_{\rm AGN} = \frac{L_{\rm  AGN}}{L_{\rm  AGN} + L_{\rm  galaxy}}\; ,
\end{equation}
where $L_{\rm  AGN}$ and $L_{\rm  galaxy}$ are integrated from the UV- to the FIR regime ($0.1$--$7175.6$\,\micron).
The parameter $f_{\rm AGN}$ was set to vary in a grid of values: $0.0, 0.1, 0.3, 0.5, 0.7$, and $0.99$.
In the latest version of \texttt{CIGALE} \citep{Yang2022ApJ...927..192Y}, the modelling of the radio emission is separated into two components: one originating from star formation and one from the AGN. The former is derived from the IR/radio correlation parameter ($q_{\mathrm{IR}}$, \citealt{Helou1985}; \citealt{Condon2002AJ....124..675C}; \citealt{Yun2001}), for which we adopted input values of 2.2, 2.5, and 2.8. The latter is computed via the radio-loudness parameter $R_{\mathrm{AGN}} = L_{\nu, \, \mathrm{5\,GHz}} / L_{\nu, \, 2500\,\text{\AA}}$, where $L_{\nu, \, \mathrm{5\,GHz}}$ and $L_{\nu, \, 2500\,\text{\AA}}$ are the monochromatic AGN rest-frame luminosities at $5\,\mathrm{GHz}$ and $2500\,\text{\AA}$, respectively. For sources with radio flux ratio $r\approx1$ (see \cref{fig:radioflux_ratios}), we adopted $R_{\rm AGN}=10, 100, 1000$, while for the remaining sources $R_{\rm AGN}=0.0, 0.5, 1.0, 5.0, 10.0$. For the remaining 35/88 sources without coverage in the ILT LOFAR observations, we have no prior information on the nature of the radio emission. Therefore, we repeated the SED-fitting procedure allowing the radio-loudness parameter to vary over the full range $R_{\rm AGN}=0.0$--$1000$. However, in the absence of high-resolution radio data, the constraints on these sources are more uncertain, and no estimates of the radio brightness temperature can be obtained.
In the \texttt{CIGALE} setup, we included a 10\% uncertainty on the fluxes, added in quadrature. In \cref{tab:cigale_params} we summarise the input parameters adopted in our analysis.

\begin{table*}
\centering
\caption{\texttt{CIGALE} modules and relative parameters used in this work. Parameters not explicitly listed are fixed to the default values set by \texttt{CIGALE}.}
\small
\begin{tabular}{llll}
\hline \hline
\noalign{\vskip 2pt}
\text{Module} & \text{Parameter} & \text{Description} & \text{Values} \\
\hline
\noalign{\vskip 2pt}
SFH (Delayed) & 
\begin{tabular}[t]{@{}l@{}}$\tau$\\$t$\\$t_{\rm bq}$\\ $r_{\rm SFR}$\end{tabular} & 
\begin{tabular}[t]{@{}l@{}}Stellar e-folding time\\Stellar age\\Burst/quench age \\ Ratio of SFR after/before $t_{\rm bq}$\end{tabular} & 
\begin{tabular}[t]{@{}l@{}} 500, 1000, 5000 Myr\\500, 1000, 3000, 5000 Myr\\ 1, 10, 100, 300 Myr \\ 0.1, 1, 10 \end{tabular}  \\
\hline
\noalign{\vskip 2pt}
\begin{tabular}[l]{@{}l@{}}Dust attenuation \\ (mod. \citealt{Charlot2000})\end{tabular} &
\begin{tabular}[l]{@{}l@{}}  $A_{V,\, \rm ISM}$ \\ Slope BC   \end{tabular} &
\begin{tabular}[l]{@{}l@{}}  ISM \textit{V}-band attenuation\\   Power law slope of the attenuation in the birth clouds \end{tabular} &
\begin{tabular}[l]{@{}l@{}} 0.3, 2.5, 4.5, 6.5 \\$ -0.7$ \end{tabular} \\
\hline
\noalign{\vskip 2pt}
\begin{tabular}[l]{@{}l@{}}Dust emission \\ \citep{Dale2014} \end{tabular} & $\alpha_{\rm d}$ & Slope & 1, 1.5, 2, 3, 4 \\
\hline
\noalign{\vskip 2pt}
AGN \citep{Fritz2006} & 
\begin{tabular}[t]{@{}l@{}}$\tau_{\rm AGN}$\\$\Psi$\\$f_{\rm AGN}$\\$\lambda_{\rm fracAGN}$\end{tabular} & 
\begin{tabular}[t]{@{}l@{}}Optical depth\\Angle between equator and line of sight\\AGN fraction \\ Wavelength range to compute $f_{\rm AGN}$\end{tabular} & 
\begin{tabular}[t]{@{}l@{}} 1.0\\0.001, 50.1, 89.99\,deg\\0.0, 0.1, 0.3, 0.5, 0.7, 0.99 \\ $0.1$--$7175.6$\,\micron \end{tabular} \\
\hline
\noalign{\vskip 2pt}
Radio & 
\begin{tabular}[t]{@{}l@{}}$q_{\rm IR}$\\$R_{\rm AGN}^{\dagger}$\\ $\alpha_{\rm SF}$ \\ $\alpha_{\rm AGN}$ \end{tabular}  & 
\begin{tabular}[t]{@{}l@{}}IR/radio correlation\\Radio-loudness parameter  \\ Radio spectral slope for SF \\ Radio spectral slope for AGN\end{tabular} & 
\begin{tabular}[t]{@{}l@{}}2.2, 2.5, 2.8\\0.0, 0.5, 1, 5, 10 / 10, 100, 1000\\$-0.7$ \\$-0.8$ \end{tabular} \\

\hline
\noalign{\vskip 2pt}
Redshift & \begin{tabular}[t]{@{}l@{}}$z$\end{tabular} & \dots & $0.0$--$6.0$ (step$\,=0.2$)\\
\hline
\end{tabular}
\tablefoot{$^{\dagger}$ The set of $R_{\rm AGN}$ values depends on the sample used (see \cref{sec:sed_fitting} for details).}
\label{tab:cigale_params}
\end{table*}

\section{Physical properties of individual sources}\label{sec:results}

In this section, we describe the main physical properties derived from the SED fitting.
We stress that, due to the reduced photometric coverage, these properties are subject to large uncertainties. This holds in particular for sources without counterparts in the FIR. For this reason, when presenting and discussing our results, we primarily focus on sources with at least one detection in the SPIRE or SCUBA-2 bands, where relevant, while the results for the remaining objects are used mainly as indicative reference values.
The resulting best-fit SEDs for the objects with at least one FIR detection are reported in \cref{app:1}, for the 53/88 sources covered by LOFAR ILT high-resolution map, while the results for the remaining 35/88 targets are shown in \cref{app:2}.
The catalogue containing the physical properties described in this section is made publicly available as additional material for this paper.

\begin{figure}
     \centering
    \includegraphics[width=\linewidth]{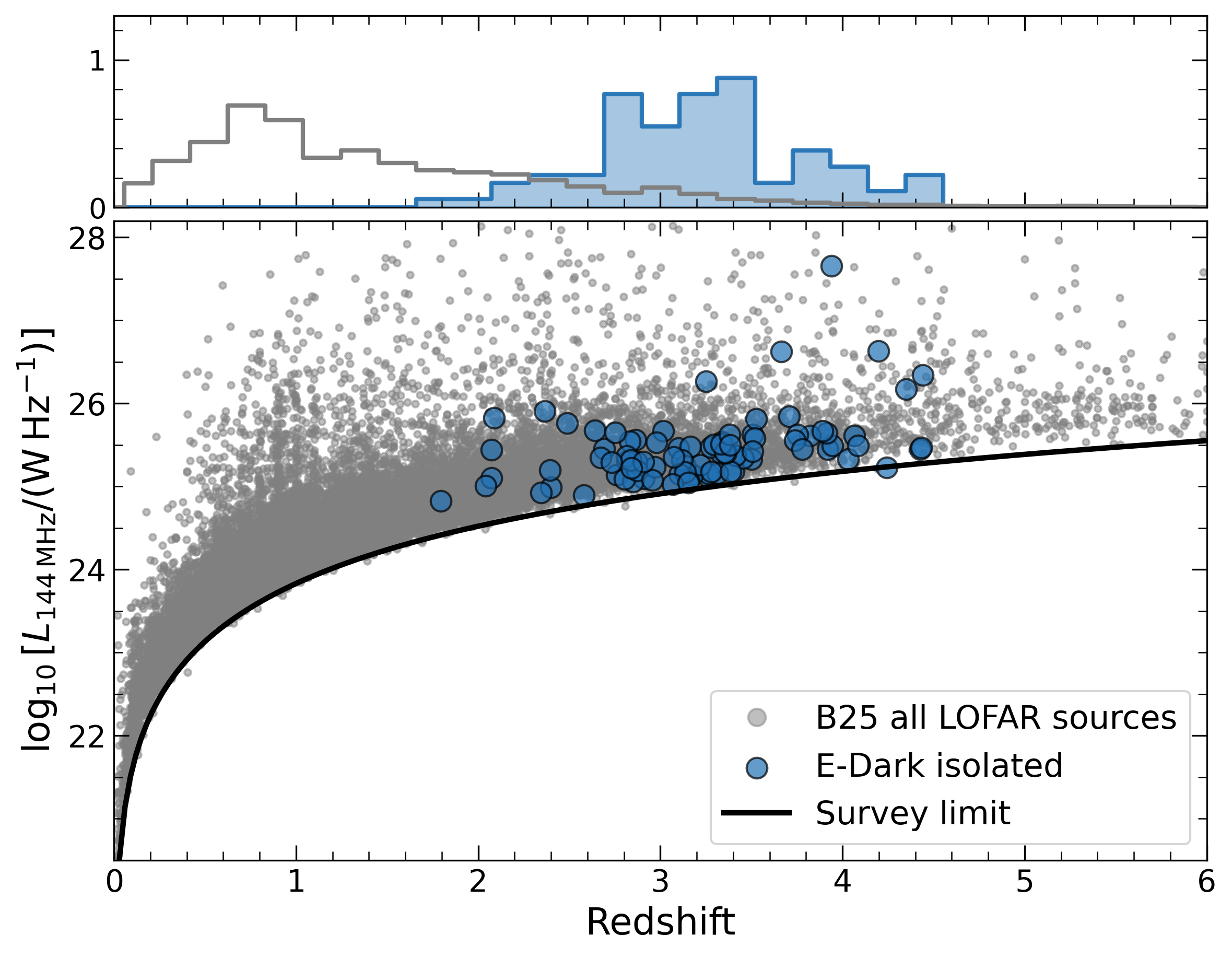}
    \caption{Radio luminosity of the 88 isolated E-dark sources (blue) as a function of redshift. In grey, we show the parent sample of LOFAR sources presented in \citetalias{Bisigello2025b}. Distributions plotted in the upper panel are normalised to the total number of sources in each sample.}
    \label{fig:z_distr}   
\end{figure}

In \cref{fig:z_distr}, we show the photometric redshift and 144\,MHz luminosity distributions for the individual sources in our sample of isolated E-dark sources. For comparison, we also report the redshift values for the entire parent sample of LOFAR sources described in \citetalias{Bisigello2025b}. Our sources span a luminosity range of $24.82\leq \log_{10}[L_{\rm 144\,MHz}/(\rm W \, Hz^{-1})]\leq 27.65$. 

The redshift distribution spans $1.8 \lesssim z \lesssim 4.4$, with a mean weighted value of $\langle z \rangle_{}=  3.4$ and a weighted standard deviation of $\sigma_{ z}=0.6$.
 All the objects of our sample are fitted with a reduced $\chi^2<1.1$. Nine sources belong to the high-$z$ tail of the distribution, with a photometric redshift $>4.0$.
The derived photometric redshifts of our sample are in good agreement with median values obtained from similar samples of optical/NIR dark radio-selected sources (e.g., $z_{\rm ph}=3.1$; \citealt{Talia2021}). 

We computed the brightness temperature as

\begin{equation}\label{eq:tb}
T_{\rm b}=1.22 \times 10^6 \,(1+z) \,\left(\frac{\nu}{1\, \mathrm{GHz}}\right)^{-2} \,\left(\frac{\theta_{\mathrm{maj}} \theta_{\mathrm{min}}}{1 \operatorname{arcsec}^2}\right)^{-1} \, \left(\frac{S_{\rm T}}{1\, \mathrm{Jy}}\right) \mathrm{~K}\,,
\end{equation}
where $\nu$ is the observed frequency (here 144\,MHz) and $S_{\rm T}$ is the corresponding total flux density. $\theta_{\rm min}$ and $\theta_{\rm maj}$ are the deconvolved major and minor axes of the fitted elliptical Gaussian components. The maximum brightness temperature attainable by a starburst galaxy, $T_{\rm b}^{\rm  SF}$, is given by \cite{Condon1991} as

\begin{equation}\label{eq:tb_threshold}
T_{\rm b}^{\mathrm{SF}} \leq T_e \,\left[1+10 \,\left(\frac{v}{1 \,\mathrm{GHz}}\right)^{0.1-\alpha}\right].
\end{equation}

\begin{figure}
     \centering
    \includegraphics[width=\linewidth]{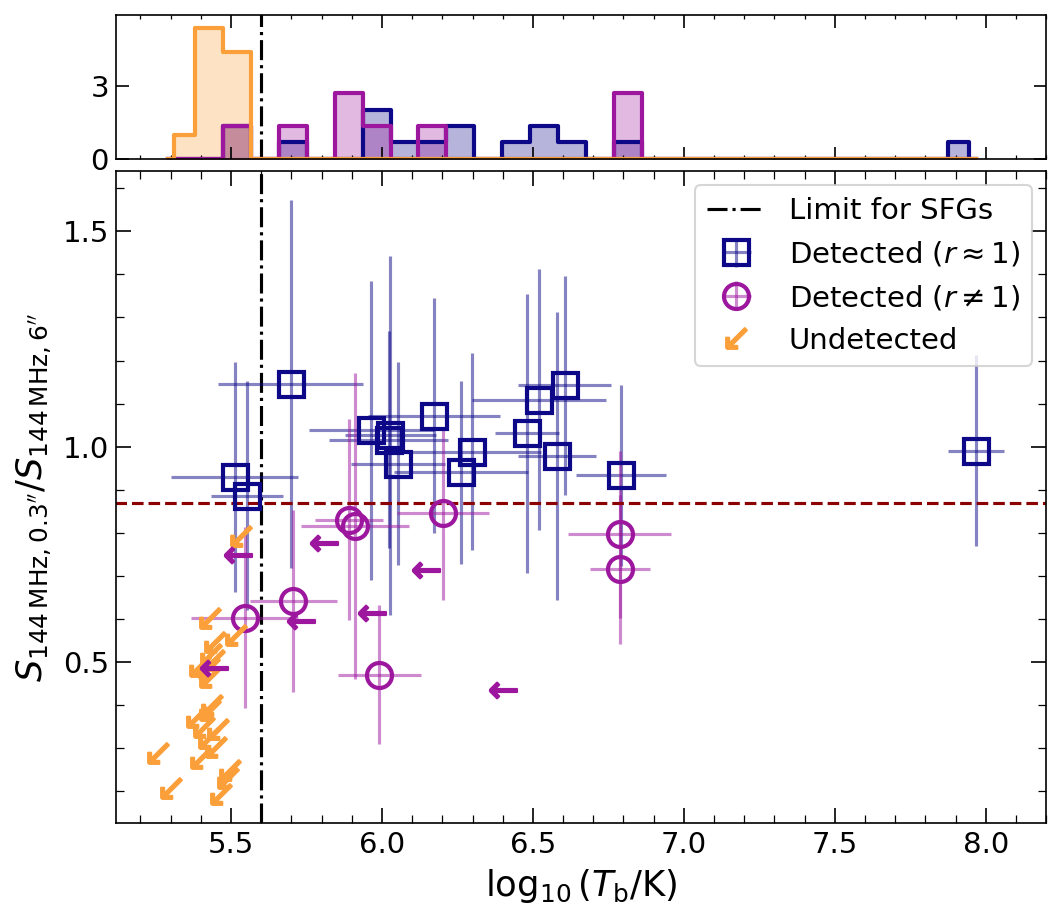}
    \caption{Ratio between 144\,MHz LOFAR fluxes at high (0\arcsecf3) and low (6\arcsec) resolution vs the brightness temperature. Vertical dash-dotted line represents the threshold adopted to distinguish between radio emission originating from star formation and AGN processes. Arrows mark $3\sigma$ upper limits for sources with either no detection in the LOFAR ILT map at 0\arcsecf3 or with an unreliable measure of the Gaussian components.}
    \label{fig:logt_distr}   
\end{figure}

In the above expression, $T_{\rm e}$ is the thermal electron temperature, here assumed to be approximately $10^4$\,K, and $\alpha$ is the radio spectral index. In \cref{fig:logt_distr}, we report the flux ratio $r$ as a function of the inferred $T_{\rm b}$. $T_{\rm b}^{\mathrm{SF}}$ is computed the spectral index value $\alpha=-0.8$, corresponding to $\logten({T_{\rm b}^{\rm SF}}/\rm K)=5.6$. We also report $3\sigma$ upper limits for sources with either no detection in the LOFAR ILT map at $0\arcsecf3$ or with an unreliable fit of the Gaussian component to the brightness distribution.
All the \num{22} sources not detected at 0\arcsecf3 (highlighted in yellow in \cref{fig:logt_distr}) show $\logten( T_{\rm b}/\rm K )<5.6$, suggesting that the radio emission in these galaxies is likely to be dominated by star formation processes. Of the \num{16} sources with $r\approx 1$, all but three have $\logten(T_{\rm b}/\rm K)>5.6$, consistent, within their uncertainties, with AGN-dominated radio emission. Of the remaining three sources, two lie below the threshold but are compatible with being above it within their error, while the third lies slightly above the threshold but is also compatible with being below it once uncertainties are taken into account. For the remaining \num{15} objects detected with $r \neq 1$, seven have upper limits on the brightness temperature, two of which lie below $\logten\left( T_{\rm b}/\rm K \right)= 5.6$, while all the remaining sources but two, have $\logten\left( T_{\rm b}/\rm K \right) > 5.6$. The two exceptions lie on opposite sides of the threshold, but are consistent with crossing it within their uncertainties. This result suggests that the radio emission in this sub-sample may originate from both AGN activity and star formation.

To estimate the possible radio excess, we derived the IR/radio correlation (IRRC) parameter ($q_{\rm IR}$), which is computed as (\citealt{Bell2003}; \citealt{Ivison2010a,Ivison2010b}; \citealt{Sargent2010})

\begin{equation}\label{eq:firrc}
q_{\mathrm{IR}}
 = \logten\!\left(
   \frac{L_{\mathrm{IR}}}{3.75\expo{12}\,{\rm W\,Hz^{-1}}}
   \right)
 - \logten\!\left(
   \frac{L_{1.4\,{\rm GHz}}}{\rm W\,Hz^{-1}}
   \right).
\end{equation}
In the above expression $L_{\rm 1.4 \, GHz}$ is the rest-frame total (i.e., including possible AGN emission) radio luminosity, computed from the 144\,MHz flux density and the photometric redshift, assuming a power-law spectrum for the radio emission $S_{\nu}\propto\nu^{\alpha}$, where $\alpha = -0.7$ \citep{Novak2017}.
$L_{\rm IR}$ is the rest-frame FIR luminosity originated by star formation and defined in the range 8--1000\,$\micron$.
\begin{figure*}
     \centering
    \includegraphics[width=\linewidth]{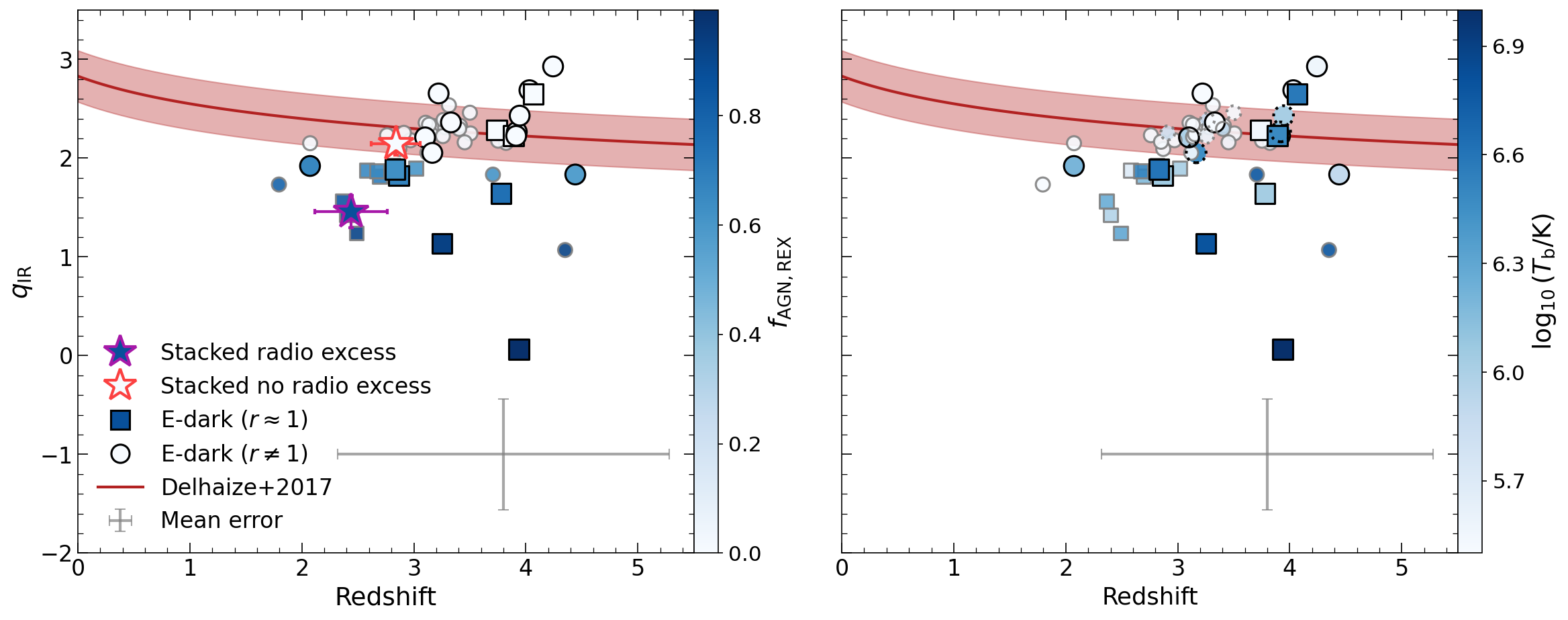}
    \caption{\emph{Left}: IRRC vs. redshift, colour-coded by AGN fraction values computed as described in \cref{sec:results}. Symbols are the same as \cref{fig:logt_distr}. Objects with at least one SPIRE detection are marked with larger symbols, while sources without SPIRE detections are shown with smaller symbols. The solid red curve represents the \cite{Delhaize2017} IRRC with its intrinsic $1\sigma$ scatter (red shaded area). The grey cross represents the mean error including all sources.
    The results for the stacked samples with and without radio excess, presented in \cref{sec:stacking}, are shown as purple- and red-outlined stars, respectively.   
    \emph{Right}: same as the left panel, but colour-coded by the brightness temperature. Upper limits in $T_{\rm b}$ are indicated by symbols outlined with a dotted line. For clarity, we only show upper limits for sources with a detection in the LOFAR ILT high-resolution image, given that all non-detections have $\log_{10} (T_\mathrm{b}/\mathrm{K})<5.6$.}
    \label{fig:irrc}   
\end{figure*}
The AGN fraction inferred from the SED fitting is subject to large uncertainties due to the poor photometric coverage in the MIR. Therefore, we follow \citetalias{Gentile2025} and estimate an AGN fraction according to the distance of our sources from the IRRC, hereafter denoted as 
$f_{\rm AGN,\, REX}$ to indicate the radio excess. For this purpose, we adopted the parametrisation of \cite{Delhaize2017}: 

\begin{equation}
    q_{\rm IR} = (2.88\pm0.03)\,(1+z)^{-0.19\pm0.01},
\end{equation}
from which we computed the AGN fraction, defined as:

\begin{equation}
    f_{\rm AGN,\, REX} = 10^{q-\bar{q}(z)}.
\end{equation}

In the above expression, $q$ is the IRRC value measured for each galaxy in our sample, while $\bar{q}(z)$ represents the value expected at the same redshift according to the relation derived by \citet{Delhaize2017}. Since this relation is characterised by an intrinsic scatter of 0.26, we assign $f_{\rm AGN,\,REX}=0$ to all sources with $q - \bar{q}(z) < 0.26$, corresponding to sources consistent within $1\sigma$ with the IRRC expected for SFGs.

\begin{figure}
    \centering
    \includegraphics[width=0.9\linewidth]{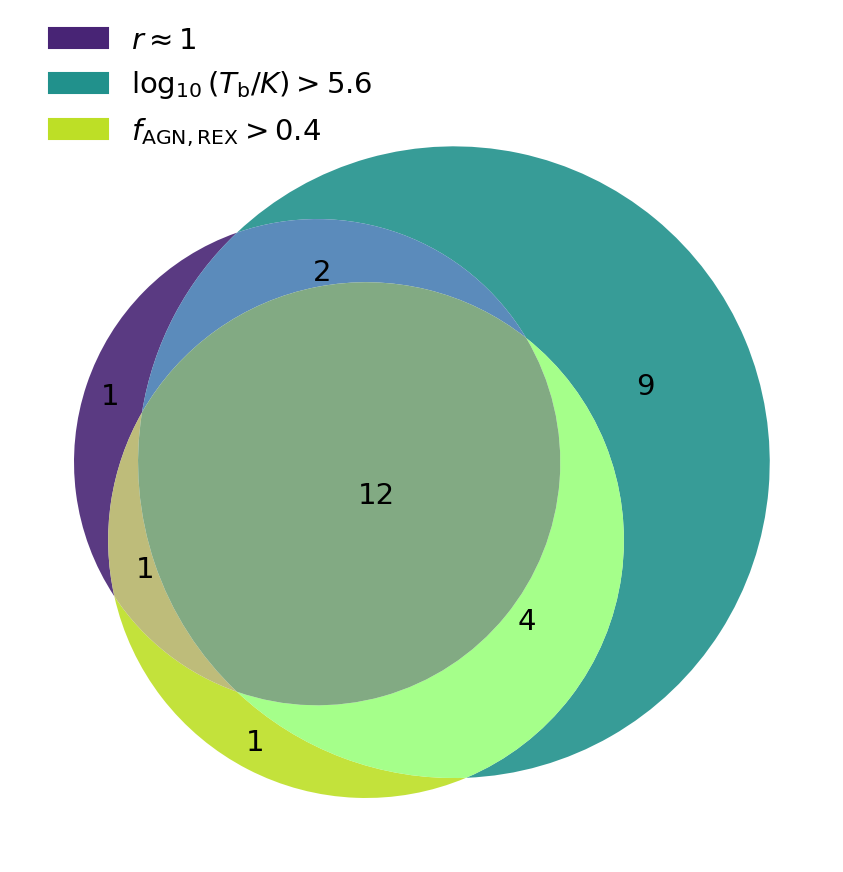}
    \caption{Venn diagram showing the sources overlapping different criteria for identifying an AGN in the radio band: excess with respect to the IRRC, flux ratio between LOFAR high and low angular resolution observations, and brightness temperature.}
    \label{fig:venn}
\end{figure}

\Cref{fig:irrc} shows the IRRC parameter for our sample as a function of the redshift and colour-coded by the AGN fraction (left panel) and $T_{\rm b}$ (right panel) values.
About 34\% of the sample shows an excess with respect to the IRRC, corresponding to $f_{\rm AGN,\, REX}>0.4$. A similar fraction is found among the 35 sources without LOFAR ILT coverage (see \cref{fig:irrc_un}), with 12 out of 35 galaxies displaying a radio excess. Focusing on the subsample of 53/88 sources with LOFAR ILT coverage, the majority of radio-excess galaxies are detected at 0\arcsecf3 at 144\,MHz, with only a few exceptions discussed below. Additionally they have $r\approx 1$ and/or $\log_{10}(T_{\rm b}/K)>5.6$. Only two sources with radio excess have $\log_{10}(T_{\rm b}/K)<5.6$ and/or have no detection in the 0\arcsecf3 LOFAR image. All the remaining sources undetected in the LOFAR ILT map are located within the IRRC. A subset of galaxies (\num{14}) lying on the IRRC is detected at high angular resolution, indicating the presence of a compact radio component, despite the absence of a radio excess.
Among these sources, nine objects either have an upper limit on the brightness temperature or exhibit $\log_{10}(T_{\rm b}/K)<5.6$, values that can be attributed to radio emission associated with star formation.
The remaining five sources are all detected at high angular resolution and have $\log_{10}(T_{\rm b}/K)>5.6$, with two of them also characterised by $r\approx 1$. For these sources, the detection of compact radio emission with high brightness temperature occurs despite their location on the IRRC.
In \cref{fig:venn} we show a Venn diagram illustrating the intersection between sources with $r \approx 1$, $\log_{10}(T_{\rm b}/K)>5.6$, and radio excess.

\begin{figure*}
     \centering
    \includegraphics[width=\linewidth]{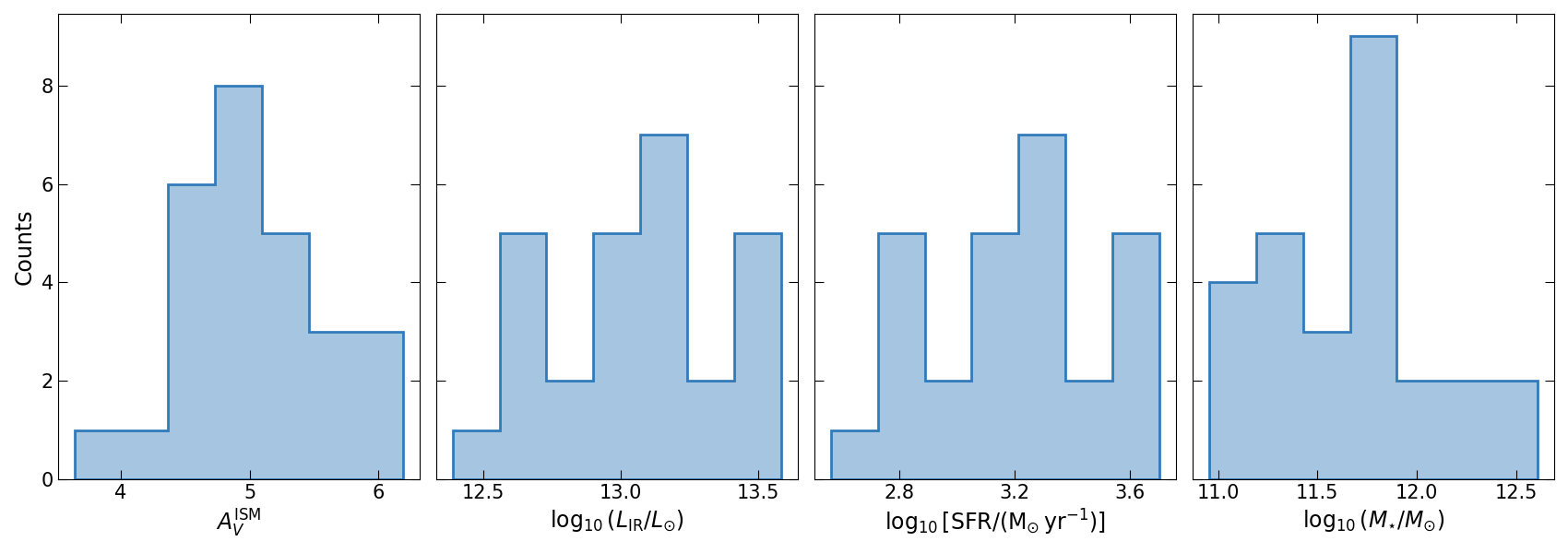}
    \caption{Distributions of the physical properties of the 26 isolated E-dark sources with at least one photometric detection in the FIR. From left to right: dust extinction of the ISM, IR luminosity, star formation rate, and stellar mass.}
    \label{fig:prop_distr}   
\end{figure*}

For the \num{26} sources with at least one FIR (SPIRE and/or SCUBA-2) detection, in \cref{fig:prop_distr} we report the distribution of the dust attenuation for the ISM, the IR luminosity of the dust component integrated in the range 8--1000\,$\micron$, the SFR derived from this latter quantity by using the standard calibration of \cite{Kennicutt2012AR}, and the stellar mass. The high obscuration values ($\langle A_V^{\rm ISM} \rangle=5.0$, $\sigma(A_V^\mathrm{ISM})=0.6$) suggest high dust content, further supported by the inferred IR luminosity ($\langle \log_{10}{(L_{\rm IR}/L_{\odot})}\rangle_{}=13.2$, $\sigma[\logten(L_\mathrm{IR}/L_\odot)]=0.3$), which places the sample in the ultra luminous infrared galaxies (ULIRGs) regime. This subset of FIR-detected isolated E-dark galaxies appears substantially more dust-obscured and massive compared to the HST-to-IRAC extremely red objects (HIERO; $\HE-{\rm IRAC2}> 2.25$, \citealt{Wang2016}) selected in the EDFs, which show median obscuration of $A_{V} = 2.3$ and stellar masses of $M_{\star} = 4 \times 10^{10}\, M_{\odot}$ (\citealt{Q1-SP016}).
Moreover, these sources show high SFR ($\langle \logten[{\mathrm{SFR}/(M_{\odot}\,\rm yr^{-1})]} \rangle_{} = 3.3$, $\sigma(\logten[\mathrm{SFR}/(M_{\odot}\,\rm yr^{-1})])=0.2$) and stellar masses ($\langle \log{M_{\star}/(M_{\odot})}\rangle=11.6$, $\sigma[\log{(M_{\star}/M_{\odot})}]=0.4$).

\begin{figure}
     \centering
    \includegraphics[width=\linewidth]{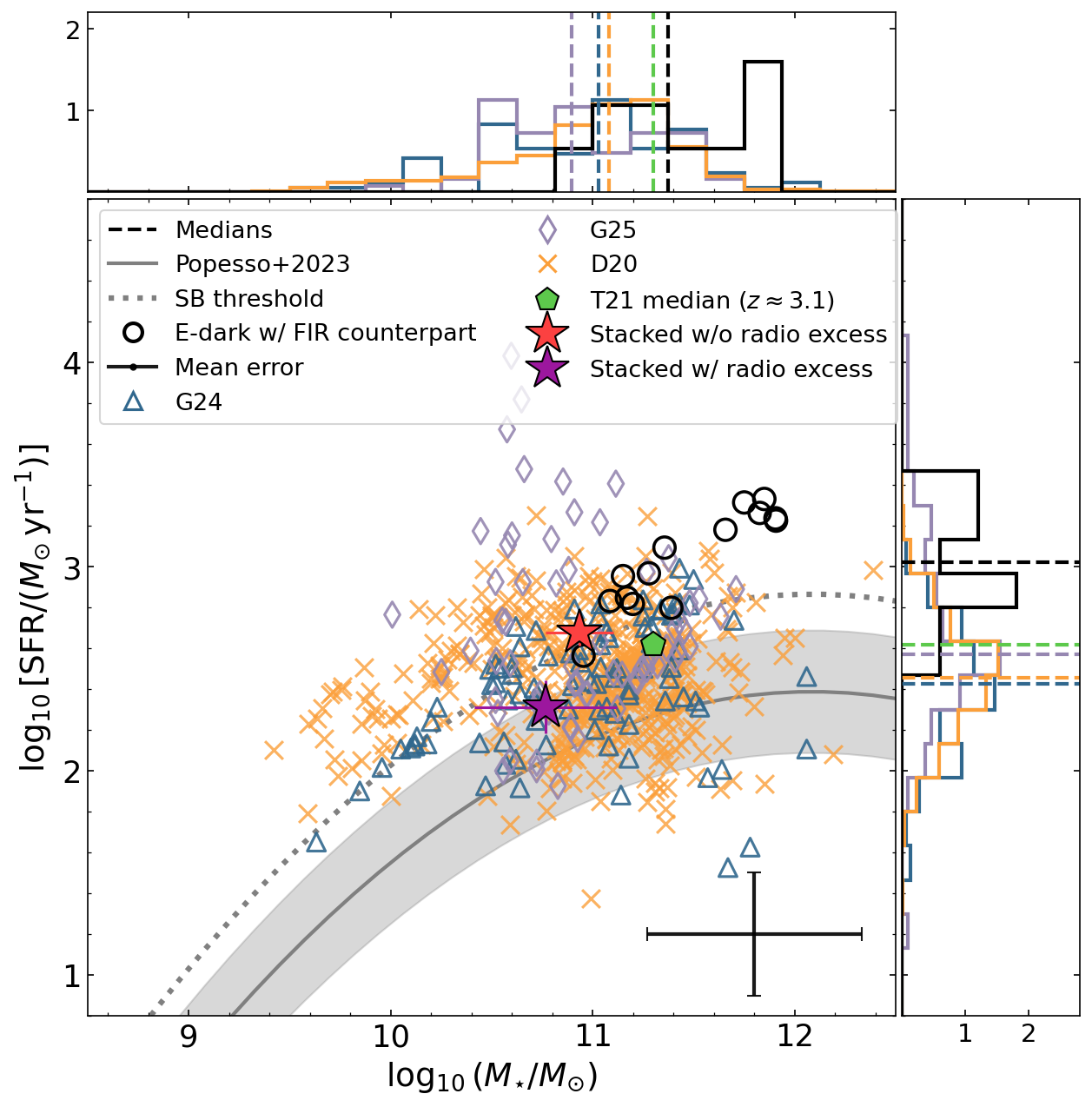}
    \caption{Star formation rate vs. stellar mass for the redshift bin $2.6 \leq  z \leq 3.6$. Empty black circles show our sample of isolated E-dark sources with at least one FIR detection. The black cross reports the mean uncertainty. Our results are compared with the MS from \citet[][at $z=3$]{Popesso2023}. The dotted line represents the threshold to consider a galaxy as a starburst (i.e., $ 3\times$ the SFR of MS galaxies from \citealt{Popesso2023}). RS-NIR dark or faint galaxies are shown as blue triangles (\citetalias{Gentile2024a}) and purple diamonds (\citetalias{Gentile2025}). For the RS-NIR dark galaxies of \citet{Talia2021}, we report the median value of the sample as a green pentagon. We also show the sample of sub-mm galaxies from \citet[][orange crosses]{Dudzeviciute2020MNRAS.494.3828D}. Purple and red stars represent the results for the stacked samples of our objects with and without radio excess, respectively (see \cref{sec:stacking}). The top and right panels show the distributions for the stellar mass and SFR for each sample, the dashed lines represent the respective median values.}

    \label{fig:main_sequence}   
\end{figure}

In \cref{fig:main_sequence} we show the the main sequence (MS; \citealt{Rodighiero2010}; \citealt{Q1-SP031}) of SFGs of \cite{Popesso2023}. We compare our results with samples obtained in literature for similar selections (\citealt{Talia2021}; \citetalias{Gentile2024a, Gentile2025}), as well as with sub-mm galaxy samples, such as the AS2UDS galaxies (\citealt{Stach2019MNRAS.487.4648S}; \citealt{Dudzeviciute2020MNRAS.494.3828D}) selected to be undetected at $K \gtrsim 25.7$.

We focus our discussion on the redshift bin $2.6\leq z <3.6$, which includes the bulk of the sources (see \cref{fig:z_distr}). We also included the MS relation by \cite{Popesso2023} computed at the mean redshift of $z=3$. 
The estimated median for the SFR and stellar mass locates the FIR-detected, isolated E-dark galaxies, approximately 0.7\,dex above the MS at $z=3$.
This subset of 26 galaxies is therefore compatible with starburst systems, spanning stellar mass ranges similar to those of the RS–NIR dark/faint galaxies presented in \cite{Talia2021}, while overlapping with the high-mass end of the RS–NIR dark samples of \citetalias{Gentile2024a} and \citetalias{Gentile2025}, as well as with the high-mass range of sub-mm–selected galaxies \citep{Dudzeviciute2020MNRAS.494.3828D}.
Similarly, our sources tend to populate the upper end of the distribution in star formation rate with respect to these samples.
By comparing the median SFR of the various samples, we find a $\Delta \log_{10}[\mathrm{SFR}/(M_{\odot}\,\rm yr^{-1})]\approx0.3$\,dex relative to \citetalias{Gentile2024a} and \cite{Talia2021}, and a $\Delta \log_{10}[\mathrm{SFR}/(M_{\odot}\,\rm yr^{-1})]\approx0.5$\,dex when compared to \citetalias{Gentile2025} and \cite{Dudzeviciute2020MNRAS.494.3828D}.
This behaviour can be explained by the fact that our subset is detected in shallower FIR observations than those adopted in these works (e.g. \citealt{Oliver2012}), favouring intrinsically brighter and more actively star-forming systems.
As a consequence, our FIR-detected galaxies represent the more extreme end of the distributions already traced by similar high-redshift samples.

\section{Stacking}\label{sec:stacking}

\begin{figure*}
     \centering
    \includegraphics[width=0.9\linewidth]{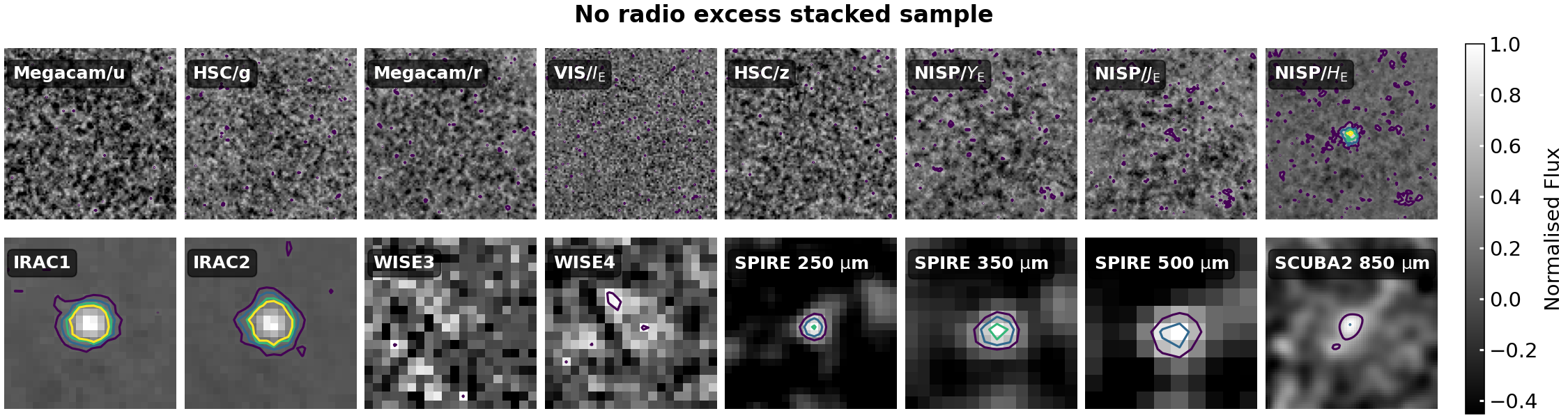}
    \includegraphics[width=0.9\linewidth]{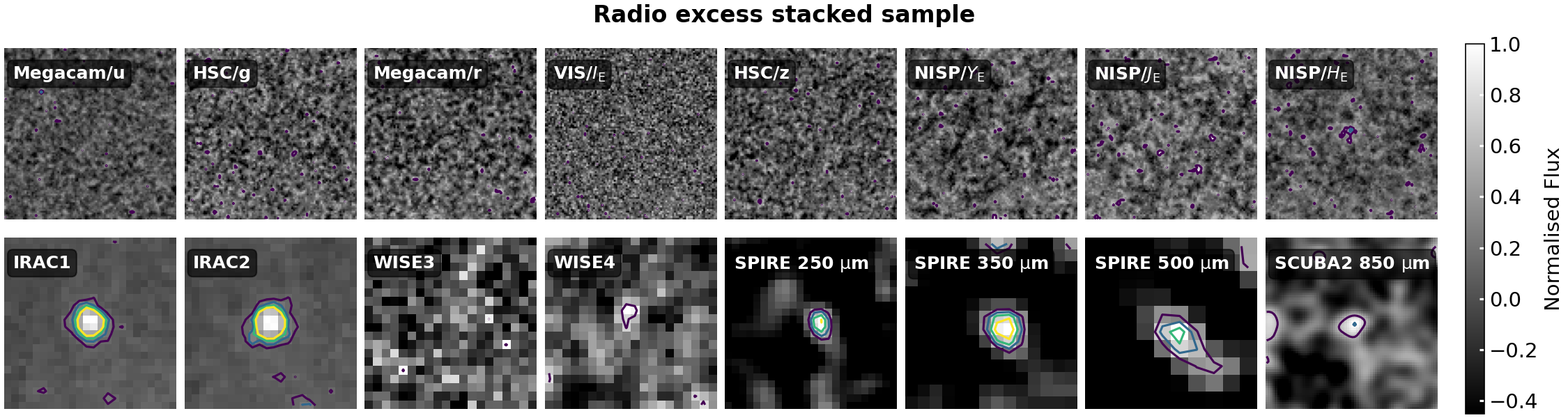}
    \caption{Images of median-stacked samples across wavebands from the UV to the FIR. The top and bottom sets of 12 images refer to our two stacked samples, respectively, without and with radio excess. Each cutout is normalised to its peak pixel value. Contours display the $3,5,7,$ and $9\sigma$ emission. Postage stamps are $10\arcsec\times10\arcsec$ for Megacam, HSC, and \Euclid, $15\arcsec\times15\arcsec$ for IRAC, $50\arcsec\times50\arcsec$ for WISE, and $150\arcsec \times 150 \arcsec$ for SPIRE and SCUBA-2.} 
    \label{fig:stacked_maps}   
\end{figure*}

To further validate and complement the results obtained for the individual sources, we performed a median stacking analysis including all 53 objects, divided according to the presence or absence of a radio excess, as identified in \cref{sec:results}. These two samples consist of \num{18} and \num{35} objects, respectively.
This approach enabled us to extract statistical information on the average properties of both subsets.
We stacked the available optical-to-FIR and low-resolution radio image cutouts, centred at the IRAC2 positions of our targets. For WISE, we adopted the unblurred unWISE images. Following \cite{Q1-SP071}, we performed stacking using the non-matched-filtered SPIRE and SCUBA-2 maps. SCUBA-2 covers, respectively, ten and nine sources among the non-radio excess and radio excess samples. The resulting median stacked optical-to-FIR images are shown in \cref{fig:stacked_maps}.

We built the median SED by extracting the photometry in each available filter as follows.
Flux densities were derived using the aperture photometry function of \texttt{PHOTUTILS}, assuming a circular aperture centred at the IRAC position. The diameter of the circular aperture was selected in correspondence to the flattening of the radial brightness profile of the source and corresponds to $1\arcsec$ for \Euclid, $6\arcsec$ for IRAC, and $9\arcsec$ for LOFAR. 
Both samples are undetected up to the NISP \JE band, and become detectable at a significance of $\approx 9\sigma$ and $\approx 3\sigma$ in \HE for the non-radio-excess and radio-excess sample, respectively.

\begin{figure*}
    \centering
   \includegraphics[width=0.48\linewidth]{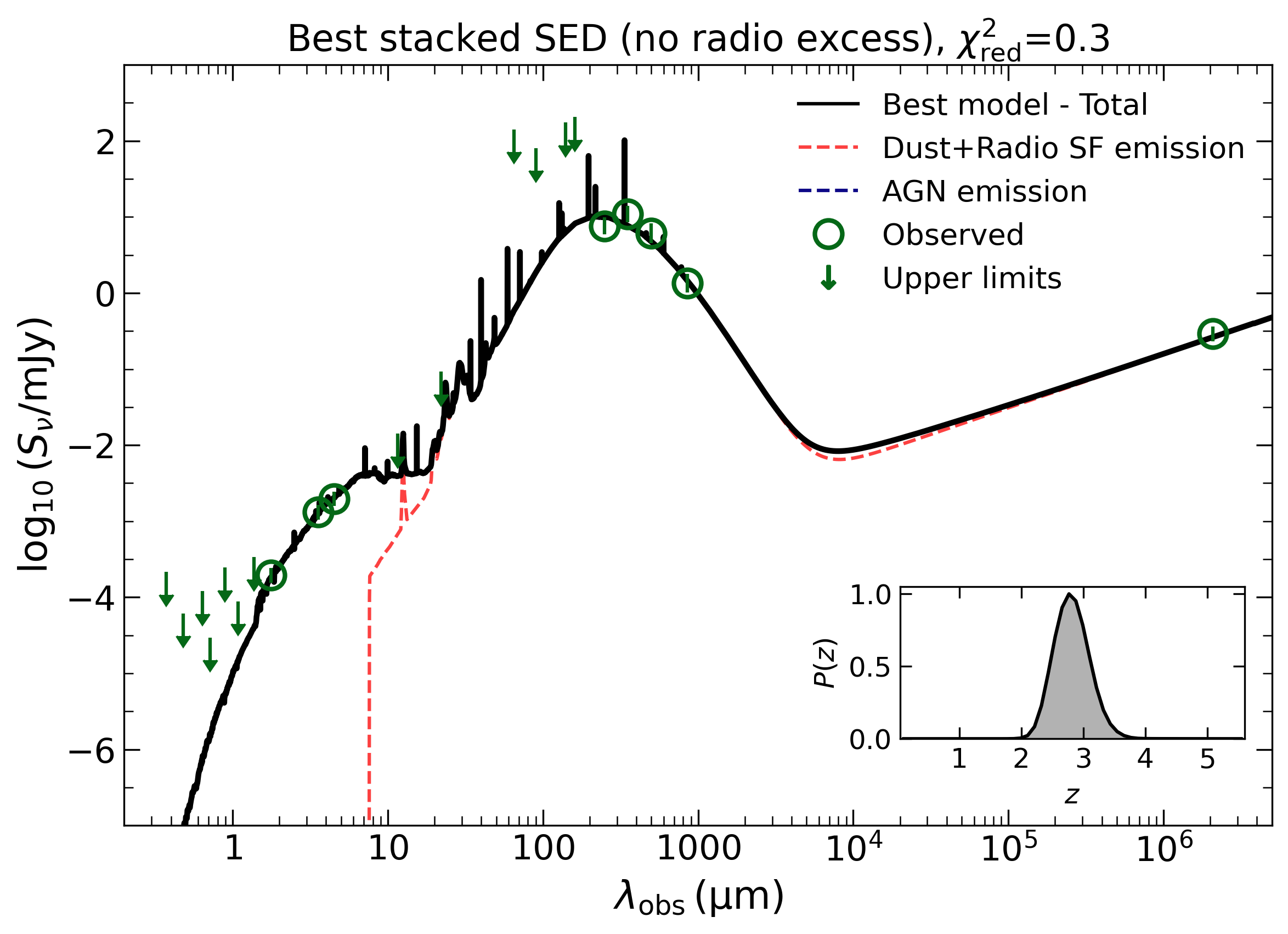}
   \includegraphics[width=0.48\linewidth]{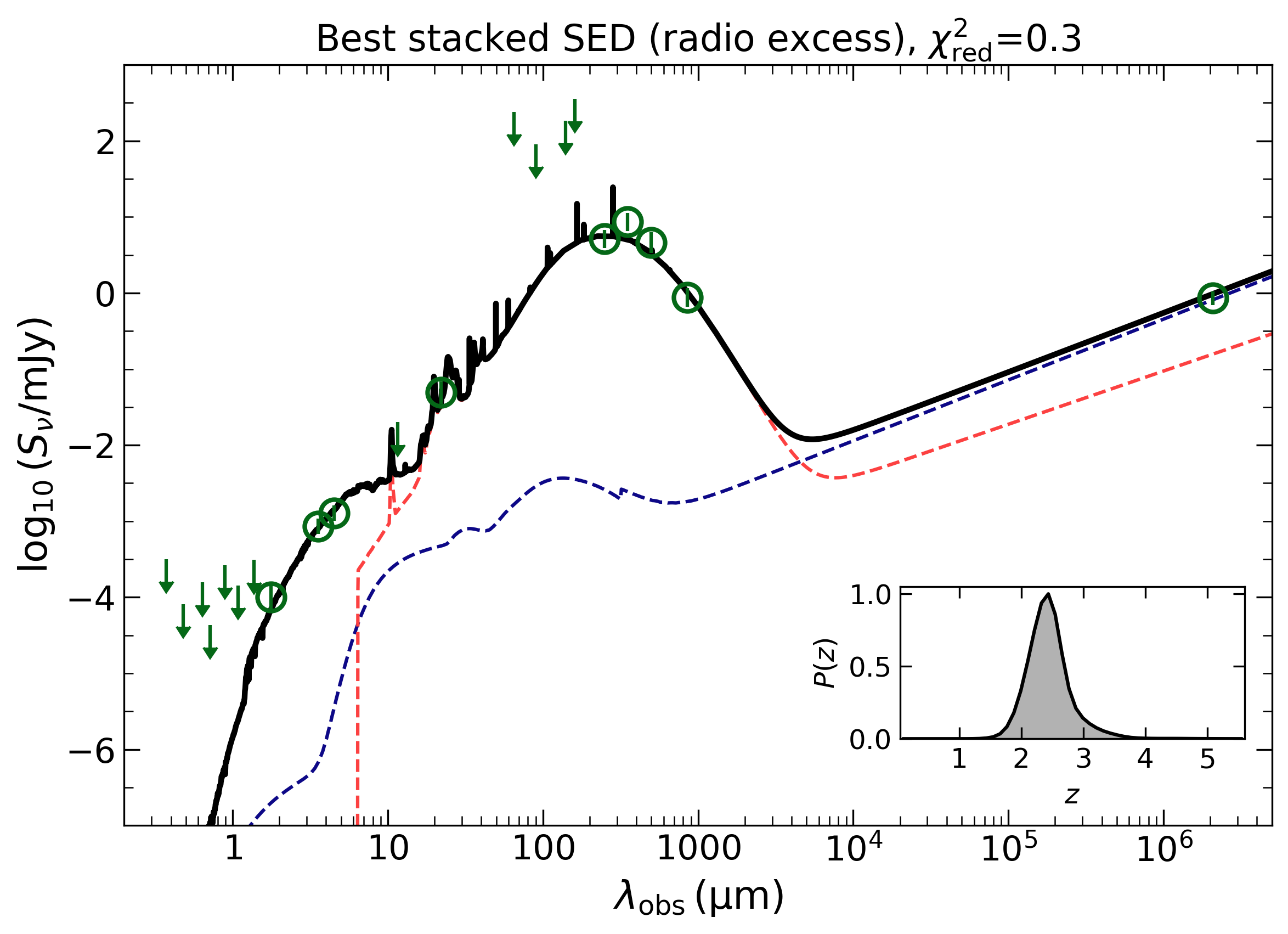}
    \caption{Best-fit SEDs from \texttt{CIGALE} for the median stacked photometry of the two samples with and without radio-excess. Empty green circles mark the photometric detections and arrows the 3$\sigma$ upper limits. The black solid line shows the total modelled emission. The dashed blue line represents the AGN contribution, including the radio component, while the dashed red line corresponds to the dust and radio emission from the host galaxy. The insert shows the probability distribution of the photometric redshift. Reported wavelengths are in the observed frame.}
  \label{fig:stacked_sed}   
\end{figure*}

We repeated the SED-fitting with the same configuration described in \cref{sec:sed_fitting}. Although the SCUBA-2 stacking does not cover the full extent of both samples, we assume that the objects included, being randomly distributed across the map, are representative of the overall populations. To verify that the SED fitting is not biased by the subset of sources within the SCUBA-2 coverage, we also repeat the analysis excluding these points as a consistency check.
In \cref{tab:median_stacking} and \cref{fig:stacked_sed}, we report the physical properties and the best-fit SED for the median stacked photometry in output from \texttt{CIGALE} for the two samples with and without radio excess.\footnote{Excluding the SCUBA-2 points from the SED fitting does not produce significant differences in the results, although the uncertainties on the derived parameters are higher. For example, we obtain with $z_{\rm ph}=3.0\pm0.6$, $\chi^2_{\rm red}=0.3$ and $z_{\rm ph}=2.7\pm0.7$, $\chi^2_{\rm red}=0.2$ for the samples without and with radio excess, respectively.}

\begin{table}
\centering
\caption{Output parameters from \texttt{CIGALE} and derived quantities.}
\label{tab:median_stacking}
\begin{tabular}{lll} 
\hline
\hline
\noalign{\vskip 2pt}
\multicolumn{3}{c}{SED-fitting results for stacking} \\ 
\hline
& No REX & REX \\
\hline
\begin{tabular}[l]{@{}c@{}}$z_{\rm ph}$ \end{tabular}  &  $2.8\pm0.2$  & $2.4\pm0.3$ \\
\begin{tabular}[l]{@{}c@{}}$\logten(M_{\rm \star} / M_{\odot})$\end{tabular} &  $10.9\pm0.2$ &  $10.7\pm0.4$   \\
\begin{tabular}[l]{@{}c@{}}$A_{V, \, \rm ISM}$ \end{tabular} &  $4.5 \pm 0.2$ & $5.1\pm0.9$ \\
\begin{tabular}[l]{@{}c@{}}$\logten[{L}_{\rm 1.4\,GHz, AGN}/({\rm W\,Hz^{-1}})]$\end{tabular}  & $[0]$ & $24.8\pm0.2$  \\
\begin{tabular}[l]{@{}c@{}}$R_{\rm AGN}$ \end{tabular} & $[0]$ & [1000]   \\
\begin{tabular}[l]{@{}c@{}} $\log_{10}(t/\rm Myr)$\end{tabular} & $2.8\pm0.1$   &  $2.9\pm0.1$ \\
\begin{tabular}[l]{@{}c@{}} $\log_{10}(\tau/\rm Myr)$ \end{tabular} & $[3.0]$  & [2.7] \\
\begin{tabular}[l]{@{}c@{}} $\alpha_{\rm d}$ \end{tabular} & $1.01\pm0.08$ & $1.4\pm0.2$ \\
\begin{tabular}[l]{@{}c@{}} $\Psi/\rm deg$ \end{tabular} & $[1\times10^{-3}]$ & $[1\times10^{-3}]$\\
\hline\noalign{\vskip 2pt}
\begin{tabular}[l]{@{}c@{}}$\logten({L}_{\rm IR}\,/L_{\odot})$\end{tabular}  & $12.6 \pm 0.1$  & $12.2 \pm 0.1$  \\
\begin{tabular}[l]{@{}c@{}}$\logten [{L}_{\rm 1.4\,GHz, SF}/(\rm W\,Hz^{-1})]$\end{tabular}  &   $24.42\pm0.01$ & $23.2\pm0.1$\\
\begin{tabular}[l]{@{}c@{}} $\logten [{\rm SFR_{\rm IR}}/( M_{\odot}$ \,yr$^{-1}$)]\end{tabular} &  $2.7\pm0.1$ &  $2.3\pm0.1$\\
\begin{tabular}[l]{@{}c@{}} $q_{\rm IR}$\end{tabular} & $2.1\pm0.1$   &    $1.2\pm0.2$\\
\begin{tabular}[l]{@{}c@{}}$f_{\rm AGN, REX}$ \end{tabular}  &  $0.3\pm0.2$   & $0.89\pm0.06$ \\

\hline
\end{tabular}
\tablefoot{Results are shown for the two stacked samples (i.e., sources with and without radio excess). From the first row: photometric redshift, stellar mass, \textit{V}-band attenuation for the ISM, AGN radio luminosity at 1.4\,GHz, radio loudness parameter, stellar age, e-folding time of the main stellar population, dust slope, and AGN viewing angle. The latter five rows report the parameters computed from the SED-fitting outputs: the IR luminosity of the galaxy computed in the range $8$--$1000$\,$\micron$, the 1.4\,GHz luminosity from star-formation processes, derived by subtracting the AGN contribution to the total radio luminosity at 1.4\,GHz, computed assuming a simple power-law spectrum $S_{\nu}\propto \nu^{\alpha}$ with $\alpha=-0.7$; the SFR inferred from the infrared luminosity exploiting the calibration of \citealt{Kennicutt2012AR}; the IR/radio correlation parameter; and the AGN fraction computed following \citetalias{Gentile2025}. For those parameters whose Bayesian estimate is not constrained from the fit (i.e., the uncertainty is higher than the parameter value), we report in square brackets the best-fit value from the input grid provided by \texttt{CIGALE}.}
\end{table}

The two samples exhibit comparable dust attenuation and show modest differences in redshift, stellar mass, and SFR, with the radio-excess sources having somewhat lower values on average.
 The stacked sample of non–radio-excess sources lies within the intrinsic scatter of the IRRC at the corresponding redshift ($z_{\rm ph} = 2.8 \pm 0.2$), corresponding to an AGN fraction of $f_{\rm AGN,\, REX}= 0.3\pm0.2$, with a $q_{\rm IR}$ value lower by 0.17\,dex than the expected value for SFGs predicted by \cite{Delhaize2017}. The radio-excess stacked sample shows a well-constrained radio excess, with an AGN fraction of $f_{\rm AGN,\,REX} = 0.87 \pm 0.05$.
For both stacked samples, the radio-loudness parameter $R_{\rm AGN}$ is poorly constrained, with \texttt{CIGALE} returning best-fit values of $R_{\rm AGN}=0$ and $R_{\rm AGN}=1000$ for the non–radio-excess and radio-excess stacked samples, respectively, where $R_{\rm AGN}=10$ is commonly adopted as the boundary between radio-quiet and radio-loud AGN \citep{Kellermann1989}.

Both stacked samples are found to lie above the locus of the star-forming MS at their corresponding photometric redshift of $z_{\rm ph}=2.8$ and $z_{\rm ph}=2.4$, for the samples without and with radio excess, respectively.
The non–radio excess stacked sample shows an offset of $\approx0.6$ dex with respect to the MS, while the radio-excess stacked sample is offset by approximately $0.4$ dex.
Given the typical intrinsic scatter of the main sequence of $0.3$ dex, the former sample can be classified as a starburst population, while the latter lies below the threshold to consider a galaxy as a starburst. At the same time, their high SFR suggest that they are not strictly representative of the bulk of main-sequence galaxies.

\section{Discussion}\label{sec:summary}

In this paper, we presented the first sample of E-dark galaxies selected in the radio band within the EDF-N field. Unlike previous studies, which mainly relied on 1.4\,GHz and 3\,GHz observations, our selection is based on deep, low-frequency (144\,MHz) LOFAR HBA data, reaching an RMS sensitivity limit of \SI{32}{\micro\jansky}\,beam$^{-1}$. Moreover, the EDF-N spans an area at least 5 times larger than previous studies, which were typically limited to smaller fields such as COSMOS and GOODS-N.

A first key difference lies in the low surface density of E-dark sources when compared to similar selections. Our selection yields \num{166} E-dark galaxies across approximately $10$\,deg$^2$, which correspond to a surface density of approximately \num{17} sources per square degree. For comparison,\citet{Enia2022ApJ...927..204E} found a surface density of $\approx 360$ sources per deg$^2$, while the RS-NIR dark galaxies found in \citetalias{Gentile2024a} and \citetalias{Gentile2025} have a surface density of approximately \num{190} and \num{235} sources per $\rm deg^{2}$, respectively.

\begin{figure}
    \centering
    \includegraphics[width=\linewidth]{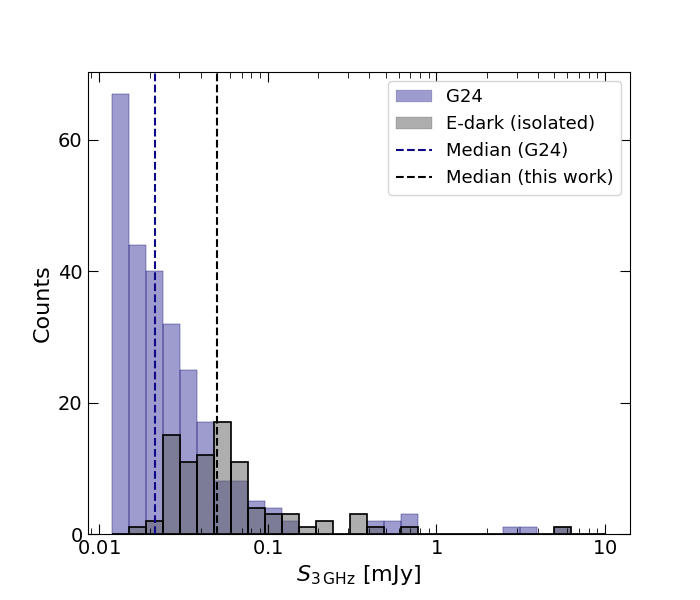}
    \caption{Comparison between the distributions of flux density at 3\,GHz of our sample of isolated E-dark sources (grey histogram) and the RS-NIR dark galaxies of \citetalias{Gentile2024a} (blue histogram). The dashed vertical lines represent the median values of the distributions.}
    \label{fig:flux_gentile}
\end{figure}

The number of sources retrieved in this paper is therefore relatively small when compared to previous samples of RS-NIR dark objects, considering that the observations exploited cover a larger area, however, a direct quantitative comparison between the samples and their number densities is hindered by the varying depths, angular resolutions, and selection criteria adopted across these studies. We therefore focus on the main differences between our selection and those adopted in previous studies. First, given that our methodology uses the \citetalias{Bisigello2025b} catalogue as a starting point, our parent sample is, by construction, limited to sources with counterparts in \textit{Spitzer}/IRAC1 and IRAC2, whose observations in the EDF-N are about one order of magnitude shallower than those conducted in COSMOS. Secondly, our selection criteria might be more stringent during the visual inspection step because we relied on lower-resolution LOFAR (1\arcsecf5) and IRAC ($\approx 2 \arcsecf$) data. At this resolution, multiple sources or contaminants may fall within the radio and infrared emission, increasing the number of objects rejected during visual inspection compared to studies based on higher-resolution data (e.g., JVLA in COSMOS). Moreover, when comparing the different radio observations, the sensitivity of LOFAR observations used in this work would reach approximately \SI{3.8}{\micro\jansky}\,beam$^{-1}$ at 3\,GHz (assuming a radio spectral index of $-0.7$), which is about a factor $1.5$ higher when compared to the VLA-COSMOS 3\,GHz Large Project survey ($1\sigma=\SI{2.3}{\micro\jansky} \, \rm beam^{-1}$) exploited in \cite{Talia2021}, \citetalias{Gentile2024a}, and \citetalias{Gentile2025} (see \cref{fig:flux_gentile}).
The relatively shallower radio sensitivity implies that our sample is limited to fewer, brighter sources, thereby excluding the fainter dark galaxies, which, as shown in \cref{fig:flux_gentile}, make up the bulk of the distribution in the sample of Rs–NIR dark sources from \citetalias{Gentile2024a}.

We analysed a sample of 88 isolated E-dark galaxies, focusing on \num{53} sources covered by high-resolution LOFAR ILT observations at 0\arcsecf3 over an area of approximately $2.5\, \rm deg \times2.5\, \rm deg$. The compact radio emission, particularly the ratio between high- and low-resolution fluxes, provided constraints on its origin and on the potential presence of AGN activity. By performing SED fitting using the \texttt{CIGALE} code, we were able to estimate the brightness temperatures $T_{\rm b}$ (or upper limits) and combine this information with the placement of these sources on the IRRC.

Our analysis shows that roughly one-third of the sample exhibits a radio excess relative to the IRRC, corresponding to an AGN fraction 
$f_{\rm AGN,\,REX}>0.4$. This fraction increases to 40\% when including AGN identified via brightness temperature. In general, the radio excess in our galaxies is associated with brightness temperatures exceeding the threshold typically used to classify objects as starbursts, whereas sources consistent with the IRRC are largely compatible with radio emission driven by star formation, with some exceptions. These exceptions likely correspond to cases with more uncertain classifications, where the radio emission is confined to sub-kpc scales and could be associated with radio-quiet AGN (e.g., \citealt{Morabito2022}). A lower abundance of radio-excess systems has been reported in similar selections based on smaller-area surveys. For instance, \citetalias{Gentile2024a} found that about $\sim$20\% of their Rs–NIR dark sample shows evidence of AGN activity when combining multiple diagnostics, such as X-ray emission and SED decomposition, while only $\sim$14\% display a radio excess when considering only the distance from the IRRC as a criterion. The higher number of radio-excess sources in our sample can be naturally explained by the wider area and relatively shallower radio coverage provided by LOFAR, which biases our selection toward the brighter end of the radio-bright and obscured population.

For 26 objects with FIR counterparts, we are able to derive estimates of the SFR and other physical properties with higher reliability. These sources represent the more extreme end of the distribution, corresponding to massive ($M_{\star}\approx10^{11}\,M_{\odot}$), obscured ($A_{V}^{\rm ISM}\approx 5$), highly star-forming ($\mathrm{ SFR}\approx 4\times 10^3\,M_{\odot}\,\mathrm{yr}^{-1}$) galaxies at intermediate redshift ($z\approx 3$), occupying the starburst locus in the SFR vs stellar mass diagram.

We performed a stacking analysis, dividing our sources into two samples according to their radio excess with respect to the IRRC. The results from the SED-fitting point out that the two samples display similar global physical properties, while differing primarily in their radio emission. The physical properties are consistent with the trends observed for individual sources, but less extreme in terms of SFR. This is expected, as these stacked sources are fainter in the FIR. Nonetheless, they generally occupy the transition region between the main sequence and starburst populations at $z=3$.
The non–radio-excess stacked sample is consistent with SFGs lying within the IRRC scatter, with no significant indications of AGN-related radio emission. In contrast, the radio-excess sample shows a clear and well-constrained radio excess, consistent with a dominant AGN contribution.
Interestingly, these results highlight that our selection efficiently identifies not only heavily obscured SFGs, but also obscured AGN at high redshift, with a significant fraction of sources exhibiting a radio excess and/or high $T_{\rm b}$ values while still hosting significant ongoing star formation. In contrast to previous selections based on smaller-area surveys, which predominantly targeted SFG, the wider-area coverage explored in this work reveals a more heterogeneous population of obscured systems, where intense star formation and significant AGN activity coexist. 

Our work, therefore, highlights the power of wide-field surveys, such as those enabled by \Euclid, in uncovering rare and heavily obscured populations. In this broader context, our analysis is a valuable pilot study to guide future selections of such populations, which offer a unique window to investigate the co-evolution of galaxies and AGN (e.g., \citealt{Mancuso2017}; \citealt{Lapi2018}).

Nonetheless, this interpretation requires further confirmation through more robust analysis, as the current limitations in photometric coverage affect the reliability of the physical properties derived via SED fitting. Future \Euclid data releases, offering deeper observations, will also improve the characterisation of these sources and allow a more accurate determination of their contribution to the cosmic SFRD. The \Euclid Data Release 1 observations are expected to include a total of 11 visits in the EDF-N, providing an additional gain in sensitivity of approximately $1.3$~mag with respect to the Q1. Moreover, the \textit{James Webb} Space Telescope, offers the possibility to follow up these galaxies through MIRI imaging, providing direct constraints on MIR emission and enabling a more reliable separation between star formation and AGN contributions in radio-selected E-dark galaxies.

\noindent
\textit{Data availability:} The table listing the 88 isolated E-dark galaxies and their physical properties is only available in electronic form at the CDS via anonymous ftp to cdsarc.u-strasbg.fr (130.79.128.5) or via \url{http://cdsweb.u-strasbg.fr/cgi-bin/qcat?J/A+A/}.

\begin{acknowledgements}

M. G. and I. P. acknowledge support from INAF under the following funding schemes: Large Grant 2022 (project "MeerKAT and LOFAR Team up: a Unique Radio Window on Galaxy/AGN co-Evolution") and Large GO 2024 (project "MeerKAT and Euclid Team up: Exploring the galaxy-halo connection at cosmic noon"). A. L. M. acknowledges support through the European Space Agency (ESA) Research Fellowship in Space Science.

We thank Smail I. for the helpful comments.
\AckEC
A complete list of acknowledgements is available on \url{https://www.euclid-ec.org/consortium/community/}.
\AckQone

\AckDatalabs
Based on data from UNIONS, a scientific collaboration using three Hawaii-based telescopes: CFHT, Pan-STARRS, Subaru \url{www.skysurvey.cc}\,. 

This publication is based on observations made with the Spitzer Space Telescope, which is operated by the Jet Propulsion Laboratory, California Institute of Technology, under a contract with NASA, and has made use of the NASA/IPAC Infrared Science Archive, which is funded by the National Aeronautics and Space Administration and operated by the California Institute of Technology. This publication makes use of the mosaics from the Spitzer Enhanced Imaging Products (SEIP, \url{https://doi.org/10.26131/IRSA433}).

The Wide-field Infrared Survey Explorer is a joint project of the University of California, Los Angeles, and the Jet Propulsion Laboratory/California Institute of Technology, funded by the National Aeronautics and Space Administration.

This paper is based on observations with AKARI, a JAXA project with the participation of ESA.

Herschel is an ESA space observatory with science instruments provided by European-led Principal Investigator consortia and with important participation from NASA.
This work is based on data from the SPIRE Point Source Catalogue (\url{https://doi.org/10.5270/esa-6gfkpzh}). The Herschel Extragalactic Legacy Project, (HELP), is a European Commission Research Executive Agency funded project under the SP1-Cooperation, Collaborative project, Small or medium-scale focused research project, FP7-SPACE-2013-1 scheme, Grant Agreement Number 607254.

This research made use of Photutils, an Astropy package for detection and photometry of astronomical sources \citep{Bradley2023}. 
\end{acknowledgements}

\bibliographystyle{aa}
\bibliography{my, Euclid, Q2, DR1} 
---------------------------------------------------------

\begin{appendix}
\twocolumn

\section{SED of individual sources}\label{app:1}

In the following, we report the best-fit SED for the 19 objects among the \num{53} LOFAR ILT-covered objects that have at least one FIR counterpart.

\begin{figure*}[b]
       \centering
    \includegraphics[width=0.9\textwidth, height=0.85\textheight]{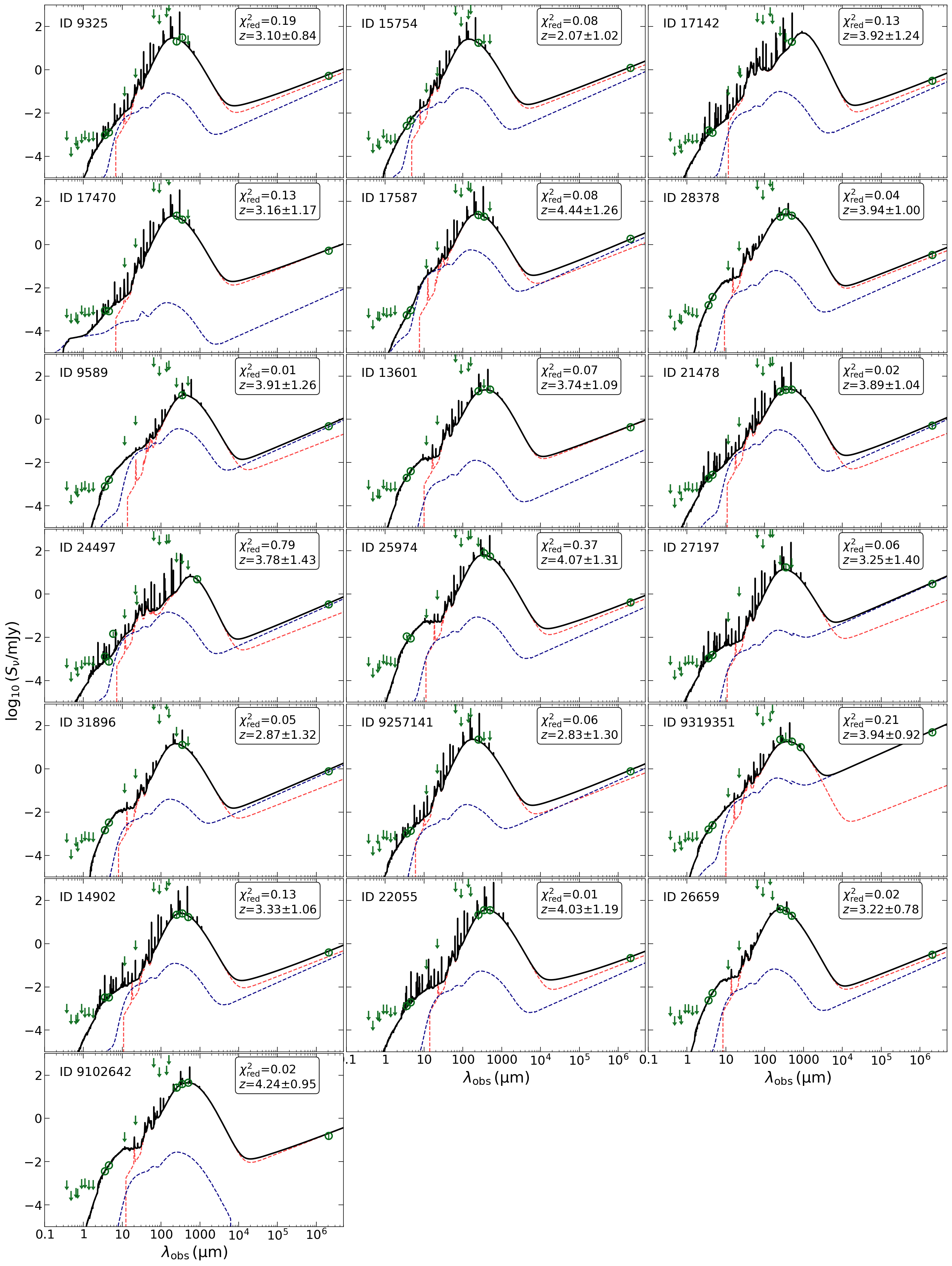}
   \caption{Best-fit SED (grey solid lines) for the 19 sources for which at least one photometric detection in the FIR is available. Symbols and lines have the same meaning as in \cref{fig:stacked_sed}. Reported wavelengths are in the observed frame.}
  \label{fig:all_seds}     
\end{figure*}

\clearpage

\section{Isolated E-dark sources with no LOFAR ILT coverage}\label{app:2}

\begin{figure*}[b]
       \centering
    \includegraphics[width=0.9\textwidth]{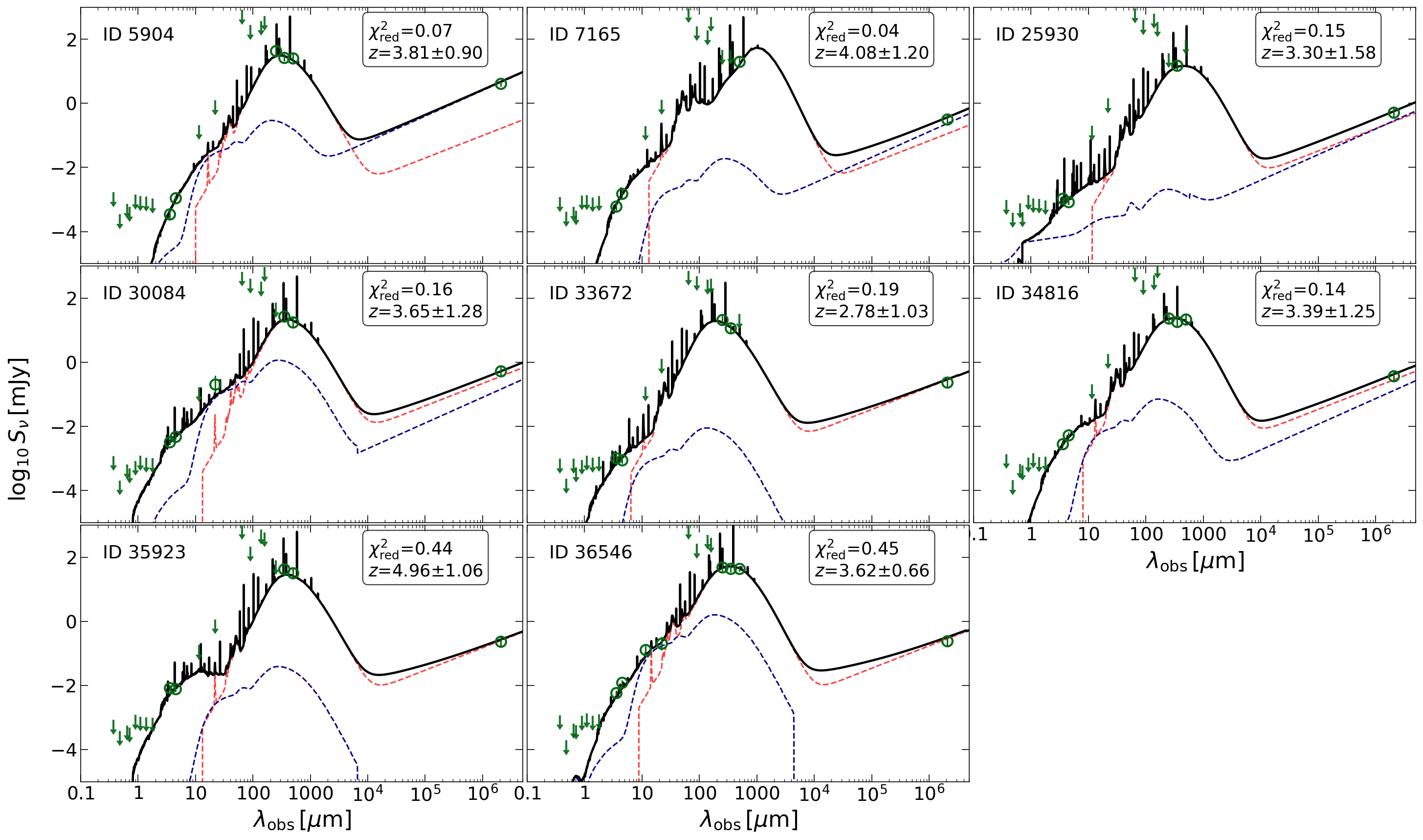}
    \caption{Same as \cref{fig:all_seds} for the sample of 8/35 sources with at least one FIR counterpart and without LOFAR ILT coverage.}
    \label{fig:all_seds_un}     
\end{figure*}

\begin{figure*}[b]
     \centering
    \includegraphics[width=0.559\linewidth]{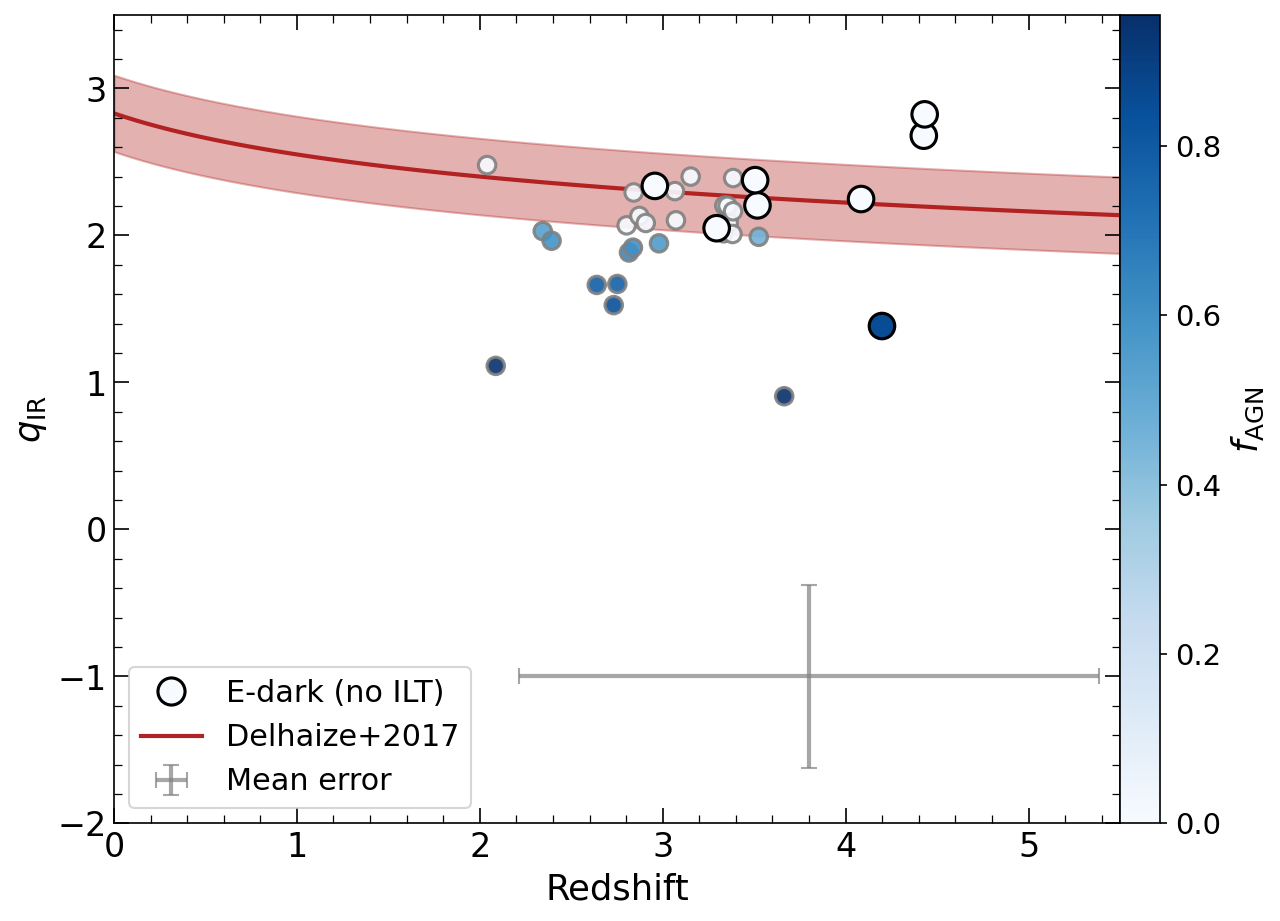}
    \caption{Same as the left panel of \cref{fig:irrc} for the 35 objects without LOFAR ILT coverage. Objects with at least one SPIRE detection are marked with larger symbols, while sources without SPIRE detections are shown with smaller symbols.}
    \label{fig:irrc_un}   
\end{figure*}

Among our sample of \num{88} isolated E-dark galaxies, \num{35} of them are not covered by the LOFAR ILT high-resolution image, and therefore no estimate of $r$ nor $T_{\rm b}$ is available. Eight sources have at least one photometric point in SPIRE and/or SCUBA-2. We repeated the SED-fitting described in \cref{sec:sed_fitting} also for this sample. Given that we have no prior knowledge on the AGN contribution to the radio emission, we allowed the radio-loudness parameter to vary as $R_{\rm AGN} = 0.0, 0.1, 1, 10, 100, 1000$. The best-fit SEDs for the sources with FIR counterparts are shown in \cref{fig:all_seds_un}, while the IRRC is shown in \cref{fig:irrc_un}.

\end{appendix}
\end{document}